\documentclass[aps,prd,onecolumn,11pt,nofootinbib,longbibliography]{revtex4-2}
\usepackage{amsmath,amssymb,amsthm,booktabs,array,graphicx,xcolor}
\usepackage{comment}
\usepackage{dsfont}
\usepackage[colorlinks,allcolors=blue]{hyperref}
\graphicspath{{figs/}}

\theoremstyle{definition}

\newcommand{\Rs}{R_{\rm s}}\newcommand{\Rb}{R_b}
\newcommand{\re}{\mathrm{Re}\,}\newcommand{\im}{\mathrm{Im}\,}

\newcommand{\lamStiff}{\lambda_{\rm stiff}}\newcommand{\lamFree}{\lambda_{\rm free}}
\newcommand{\ginf}{g_\infty}

\begin{document}

\title{Bounds on Love Numbers and Quasinormal Modes 
\\ from the Surface Impedance}

\author{Daniel McLoughlin}
\email{dmcloughlin@physics.ucla.edu}
\affiliation{Mani L.~Bhaumik Institute for Theoretical Physics, Department of Physics and Astronomy, University of California, Los Angeles, CA 90095, USA}
\author{Mikhail P.~Solon}
\email{solon@physics.ucla.edu}
\affiliation{Mani L.~Bhaumik Institute for Theoretical Physics, Department of Physics and Astronomy, University of California, Los Angeles, CA 90095, USA}

\begin{abstract}
The surface impedance of a body is fixed by its interior alone and encodes its entire tidal response.
We derive bounds on Love numbers and quasinormal modes from constraints on the impedance that follow from either of two properties of a stable body: it is lossless, meaning that it dissipates no energy, or it is passive, meaning that it never gives back more energy than it has received.
For a lossless body, every dynamical Love number is bounded from above and below, exactly in general relativity, in any renormalization scheme and at any scale.
Earlier bounds are exact for Newtonian stars but fail for compact stars, because they take the response to arbitrarily fast tides to be zero or non-negative, whereas in general relativity it is negative.
The quasinormal modes of a lossless body are also excluded from a region of the complex frequency plane set by the mass and radius.
Each known Love number enlarges the excluded region.
Conversely, the quasinormal modes bound each Love number from below, given the lower ones.
For a passive body of horizon scale, such as a black hole, the static Love number lies in a window that closes onto the vanishing black-hole value as the body's surface approaches the horizon.
\end{abstract}

\maketitle

\section{Introduction}
\label{sec:intro}

Tidal Love numbers describe how a compact body deforms in the presence of a tidal field \cite{Flanagan:2007ix,Hinderer:2007mb,Binnington:2009bb,Damour:2009vw,Chakrabarti:2013lua,Hinderer:2016eia,Steinhoff:2016rfi,Andersson:2019dwg}.
Their imprint on gravitational waves provides a direct probe of the body's internal structure.
The binary neutron-star signal GW170817 already constrained the tidal deformability of neutron stars, and with it the equation of state of dense matter \cite{LIGOScientific:2017vwq,LIGOScientific:2018hze}.
The same signal has been used to infer the frequencies of the stars' fundamental oscillation modes \cite{Pratten:2019sed}.

Static Love numbers encode the leading response to a slowly varying tide. For black holes in general relativity these are exactly zero~\cite{Binnington:2009bb,Kol:2011vg,Hui:2020xxx,Charalambous:2021mea}. This presents a test for future detectors: a nonzero static Love number for an object that otherwise resembles a black hole would signal new physics~\cite{Cardoso:2017cfl,Cardoso:2019rvt}.
The upcoming Laser Interferometer Space Antenna (LISA) \cite{LISA:2024hlh} will be particularly sensitive to the Love numbers of the massive central object in an extreme-mass-ratio inspiral \cite{Pani:2019cyc,Piovano:2022ojl}.
Quasinormal modes (QNMs), the complex frequencies at which an object rings, offer a second probe of internal structure. 
An object that mimics a black hole but rings at different frequencies would likewise signal new physics~\cite{Cardoso:2016rao,Cardoso:2019rvt}.

The space of models these measurements can test is large, which raises a basic question: which values of the Love numbers are physically allowed?
A natural tool for this question is effective field theory (EFT).
In EFT, Love numbers are Wilson coefficients that parametrize the departure from the point-particle approximation~\cite{Goldberger:2004jt,Goldberger:2005cd,Porto:2016pyg}.
This framework provides a systematic way to compute finite-size effects on the orbit and the waveform to high order~\cite{Bini:2012gu,Henry:2019xhg,Henry:2020ski,Cheung:2020sdj,Kalin:2020lmz,Bern:2020uwk,Mandal:2023lgy,Jakobsen:2023pvx}, and it gives Love numbers a precise definition.
In general relativity, the field the body produces in response to a tide cannot be separated unambiguously from the tide's own scattering off the body's gravitational field~\cite{Fang:2005qq,Gralla:2017djj}.
The EFT resolves this through matching. 
The static Love number is then defined unambiguously, while the dynamical Love numbers, which describe how the response changes with the tide's frequency, are defined only once a length scale is chosen and depend logarithmically on it, a dependence known as renormalization-group running~\cite{Charalambous:2021mea,Saketh:2023bul,Ivanov:2024sds,Katagiri:2024wbg,Caron-Huot:2025tlq,Combaluzier--Szteinsznaider:2025eoc,Kobayashi:2025vgl,Kosmopoulos:2025rfj,Solon:2026ubm,Garcia-Saenz:2026pcb}.

The EFT parametrizes the response of any body in a model-independent way, without reference to its internal structure.
For the same reason, it does not restrict the Love numbers themselves, and constraints must instead come from physical principles about the body it describes.
In particle physics, the corresponding problem is addressed by positivity bounds: causality makes amplitudes analytic, unitarity makes their imaginary parts positive, and dispersion relations turn these into inequalities on the Wilson coefficients~\cite{Adams:2006sv,Bellazzini:2020cot,Arkani-Hamed:2020blm,Caron-Huot:2020cmc,deRham:2022hpx}.
For gravity, however, this program meets well-known obstacles.
The exchange of a graviton makes the amplitude singular exactly in the forward limit that the simplest arguments rely on, and the analytic structure of scattering is intricate, even for a black hole~\cite{Caron-Huot:2022ugt,Caron-Huot:2021rmr,Correia:2025enx}.
Tidal effects enter the amplitude only at high order in $G$, together with long-range effects from the scattering of the wave off the body's own gravitational field. This scattering produces logarithms and other non-analytic terms independent of the body~\cite{Blanchet:1987wq,Goldberger:2009qd,Ivanov:2022hlo,Saketh:2023bul,Caron-Huot:2025tlq,Kosmopoulos:2025rfj,Ivanov:2026icp,Solon:2026ubm,Correia:2026utp}.

\begin{figure}[t]
\centering
\includegraphics[width=0.85\textwidth]{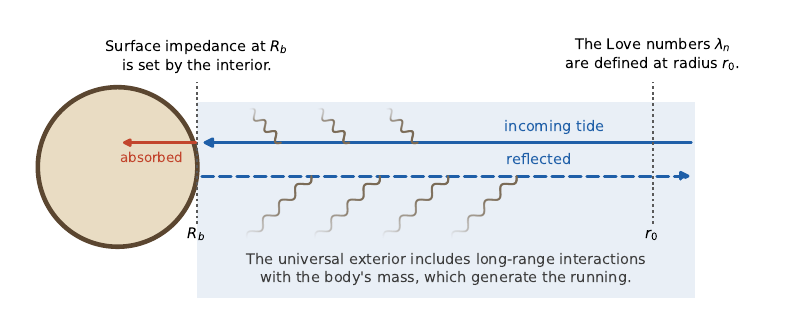}
\caption{The surface impedance, defined at the body's radius $\Rb$, is set by the interior alone.
By Birkhoff's theorem, the exterior (shaded blue) of a spherically symmetric body is the Schwarzschild geometry, the same for every body of mass $M$.
It contains the long-range interactions of the waves with the body's mass (wavy lines), which make the dynamical Love numbers run.
We bound the impedance at $\Rb$ and then use the exterior solution to translate these bounds into the Love numbers $\lambda_n$, defined far from the body at a reference radius $r_0$, in any scheme and at any scale.}
\label{fig:schematic}
\end{figure}

A more direct route bounds the tidal response itself using its causality and passivity together with an assumption about how it behaves at large frequency~\cite{Kehagias:2026opd,Apostolidis:2026mvb,Kehagias:2026nqi}, as is done for the electromagnetic response of media in Refs.~\cite{Creminelli:2024lhd,Creminelli:2025rxj}.
For Newtonian stars, whose tidal response is a sum over normal modes~\cite{Chakrabarti:2013xza,Andersson:2019ahb}, the resulting bounds hold.
In general relativity, however, they run into three difficulties. 
First, they hold at some values of the renormalization scale and fail at others because dynamical Love numbers run. 
Second, the argument assumes that the response grows more slowly than linearly at large complex frequency, which fails in general relativity.
Third, the response to an arbitrarily fast tide is assumed non-negative, whereas in general relativity it is negative.

In this paper, we derive bounds on Love numbers that hold exactly in general relativity, at any compactness, in any renormalization scheme, and at any scale.
The key is to bound the body itself through properties of its surface impedance, which is fixed by the interior alone.
The first two difficulties come from bounding the response far from the body, where it includes long-range effects such as logarithms that make the Love numbers run and a frequency-dependent dressing that spoils the response's behavior at large frequency.
Since long-range effects are fixed by the exterior equation alone and are the same for every body of a given mass, they can be separated from the body's own response; see Fig.~\ref{fig:schematic}. 
Working instead with the impedance, we place bounds at the surface before dressing with the exterior. 
The bounds then inherit the running rather than being violated by it, and the response at the surface approaches a constant at large frequency.
This removes the third difficulty: we compute the constant rather than assume its sign, and it is the Love number of an infinitely stiff surface, which is negative.

Our bounds apply to any stable body with one or both of two physical properties, each of which constrains its impedance:
\begin{itemize}
\item A \emph{lossless} body dissipates no energy.
\item A \emph{passive} body never gives back more energy than it has received.
\end{itemize}
\begin{figure}[t]
\centering
\includegraphics[width=\textwidth]{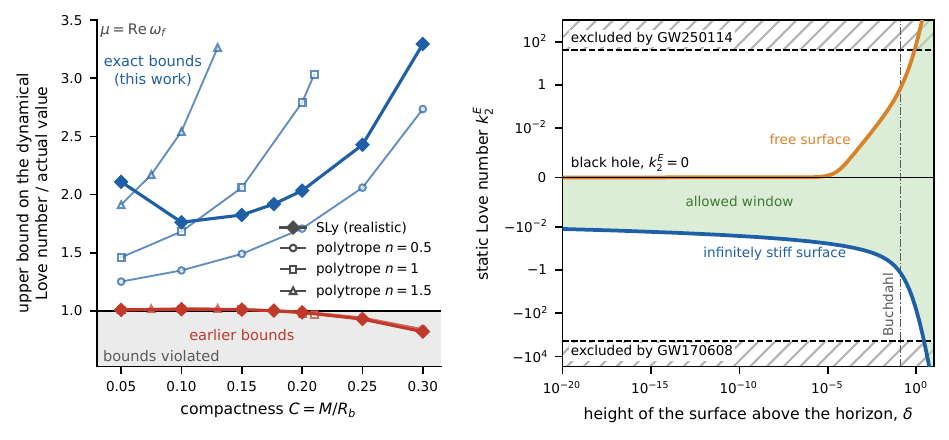}
\caption{LEFT: Upper bounds on the leading dynamical Love number $\lambda_2$ of a lossless body, applied to relativistic stars at the renormalization scale $\mu=\re\omega_f$, with $\omega_f$ the frequency of the star's fundamental mode, divided by each star's actual value, versus compactness $C=M/\Rb$.
The stars are built from four equations of state: SLy, a realistic equation of state for neutron-star matter~\cite{Douchin:2001sv}, and polytropes of index $n=0.5$, $1$ and $1.5$.
For SLy we use the piecewise-polytrope fit of Ref.~\cite{Read:2008iy}.
Each sequence stops below its maximum-mass configuration.
A ratio above one means the bound holds.
Our exact bounds (blue) hold for every star, while the earlier bounds of Refs.~\cite{Kehagias:2026nqi,Apostolidis:2026mvb} (red) exceed the actual values by up to about $2\%$ for $C<0.2$ and fall below them for $C\ge0.2$.
RIGHT: Allowed window (green) for $k^E_2$ of a non-rotating passive body, versus the height $\delta$ of its surface above the horizon, in units of the Schwarzschild radius.
Here $k^E_2=\lambda_0/(12M^5)$ is the static Love number normalized by the mass.
The upper and lower bounds (orange and blue) correspond to a free and an infinitely stiff surface; between them lies the black-hole value $k^E_2=0$ (black line).
The vertical axis is logarithmic everywhere except for $|k^E_2|\lesssim10^{-4}$, where it is linear.
The hatched regions are excluded by current bounds from GW250114~\cite{Andres-Carcasona:2025bni} and GW170608~\cite{Narikawa:2021pak}.
The vertical dash-dotted line marks the Buchdahl height $\delta=1/8$, below which no fluid star exists.}
\label{fig:love}
\end{figure}
The stars we study are lossless.
This is known in Newtonian gravity; we prove it for axial tides of relativistic perfect-fluid stars, and we check it numerically for polar tides of relativistic stars.
Black holes are passive but not lossless.
For horizon-scale bodies we assume only that they are passive.
We treat linear gravitational tides of both parities~\cite{Regge:1957td,Zerilli:1970se} on non-rotating bodies, for which the exterior is the Schwarzschild geometry, at multipole order $\ell\ge2$. 

We highlight the following results:
\begin{itemize}
\item \emph{Bounds on every dynamical Love number.}
For lossless bodies, we bound every dynamical Love number from above and below.
The bounds depend only on the static Love number, the mass and radius of the body, and the lowest resonant frequency $\omega_1$ of the body's response, which for the stars we compute lies close to the frequency of the fundamental mode.
\item \emph{Exact bounds in general relativity.}
Earlier bounds, applied as in Refs.~\cite{Kehagias:2026nqi,Apostolidis:2026mvb}, fail for compact stars, while ours hold for every star we have computed (Fig.~\ref{fig:love}, left).
Moreover, ours hold at every renormalization scale, while earlier bounds hold at some scales and fail at others (Fig.~\ref{fig:running}).

\begin{figure}[t]
\centering
\includegraphics[width=0.5401\textwidth]{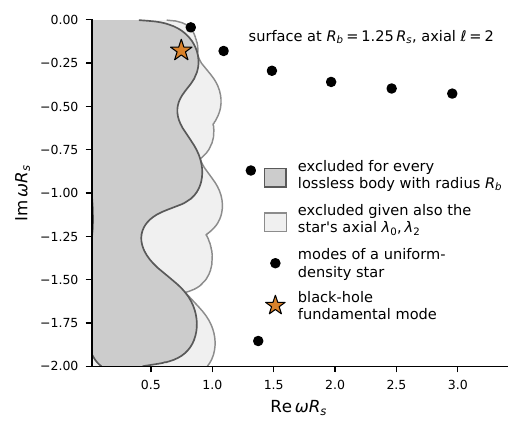}%
\includegraphics[width=0.4599\textwidth]{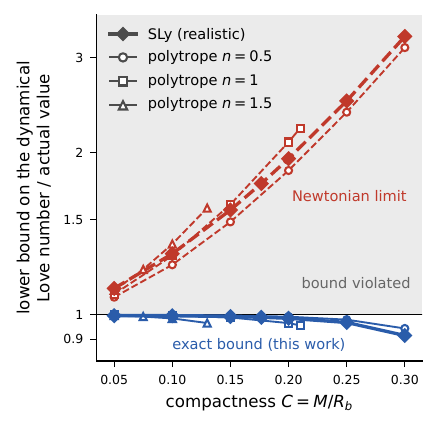}
\caption{LEFT: QNM frequencies excluded for any lossless body whose surface is at $\Rb=1.25\,\Rs$ (dark gray).
Frequencies are in units of $1/\Rs$, with $\Rs=2M$, so the region depends only on $\Rb/\Rs$.
Knowing a body's static and leading dynamical Love numbers excludes more (light gray, shown for a uniform-density star of this radius), and further Love numbers exclude more still (Fig.~\ref{fig:qnmN}).
The star's modes (dots) lie outside both regions; the least-damped one lies just outside the light gray region, within $0.01/\Rs$ of its edge.
As a reference point, we include the fundamental mode of a black hole (orange star), which lies inside the dark gray region; a black hole is not lossless. 
All modes shown are axial and quadrupolar ($\ell=2$).
RIGHT: The static Love number and the fundamental mode bound the leading dynamical Love number from below.
Shown is the lower bound on $\lambda_2$ given the static Love number $\lambda_0$ and the complex frequency of each star's fundamental mode, divided by the actual value, for the stars of Fig.~\ref{fig:love} at the scale $\mu=\re\omega_f$.
Our bound (blue) lies within $9\%$ of $\lambda_2$ for every star, while its Newtonian limit (red) fails for every star.
These are polar modes, for which we check numerically that the stars are lossless.}
\label{fig:qnm}
\end{figure}

\item \emph{A window around the black-hole value.}
For a passive body, the static Love number lies in a window (Fig.~\ref{fig:love}, right). 
Its lower bound is the Love number of an infinitely stiff surface~\cite{Cardoso:2017cfl,Chakraborty:2026qru}.
The window closes onto the black-hole value as the height $\delta$ of the surface goes to zero, and the wormholes and gravastars of Ref.~\cite{Cardoso:2017cfl} lie inside it. 
\item \emph{Bounds on quasinormal modes.}
A QNM is a frequency at which the field is purely outgoing far from the body~\cite{ChandrasekharFerrari1991,Kokkotas:1999bd,Maggio:2020jml}.
For a lossless body, the impedance is restricted at each complex frequency, which excludes QNMs from a region of the complex frequency plane that depends only on the mass and radius of the body (Fig.~\ref{fig:qnm}, left).
Knowing also the Love numbers enlarges the excluded region (see also Fig.~\ref{fig:qnmN}, right).
\item \emph{Love numbers from the fundamental mode.}
Read the other way, the same conditions bound each Love number from below, given the lower ones.
In particular, the body's mass, radius, static Love number and the complex frequency of its fundamental mode bound the leading dynamical Love number from below.
For every star we have computed, this bound lies within $9\%$ of the actual value, while its Newtonian limit fails (Fig.~\ref{fig:qnm}, right).
\end{itemize}

The rest of the paper is organized as follows.
Section~\ref{sec:surface} defines the surface impedance, states the two properties and their mathematical form, and introduces the bodies we study.
Section~\ref{sec:schemes} explains how a Love number is obtained from the impedance in any scheme and at any scale.
Section~\ref{sec:bounds} derives the window for the static Love number, including horizon-scale bodies, the bounds on the dynamical Love numbers and their test on relativistic stars, and the bound for bodies with both properties.
Section~\ref{sec:qnm} bounds the QNMs of a lossless body, also using its Love numbers, and applies the same conditions in reverse to bound each Love number from below.
Section~\ref{sec:discussion} concludes.
Appendix~\ref{app:stability} collects the consequences of stability, Appendix~\ref{app:energy} discusses the choice of energy at the surface, and Appendix~\ref{app:stars} lists the data needed to reproduce the figures.

\section{The surface impedance and two properties of the body}
\label{sec:surface}

In this section, we show that the body's tidal response is captured by the surface impedance.
We then translate the two properties, lossless and passive, into conditions on the impedance that ultimately produce our bounds.

\subsection{The exterior}

By Birkhoff's theorem the geometry outside a spherically symmetric body of mass $M$ is the Schwarzschild geometry,
\begin{equation}
\mathrm{d}s^2=-f\,\mathrm{d}t^2+\frac{\mathrm{d}r^2}{f}+r^2\mathrm{d}\Omega^2,\qquad f\equiv1-\frac{\Rs}{r},
\end{equation}
where $\Rs=2M$ is the Schwarzschild radius.
A linear gravitational perturbation of this geometry splits into multipoles $(\ell,m)$, and each multipole into two parities.
For each multipole and parity the perturbation outside the body is described by a single function $u(r)e^{-i\omega t}$, the Regge--Wheeler function~\cite{Regge:1957td} for odd-parity (axial) and the Zerilli function~\cite{Zerilli:1970se} for even-parity (polar) perturbations.
It obeys a wave equation 
\begin{equation}\label{eq:master}
\left(\frac{\mathrm{d}^2}{\mathrm{d}r_*^2}+\omega^2-V\right)u=0, \qquad r>\Rb,
\end{equation}
where $\Rb$ is the radius of the body's surface, $r_*$ is the tortoise coordinate defined by $\mathrm{d}r_*=\mathrm{d}r/f$, and the potential is
\begin{equation}
V_{\rm RW}=f\left(\frac{\ell(\ell+1)}{r^2}-\frac{3\Rs}{r^3}\right),\qquad
V_{\rm Z}=f\left(\frac{\Rs}{r^3}+\frac{2\alpha}{3r^2}+\frac{8\alpha^2(2\alpha+3)}{3(2\alpha r+3\Rs)^2}\right),
\end{equation}
with $\alpha=(\ell-1)(\ell+2)/2$~\cite{Hui:2020xxx}.
We use two solutions of Eq.~\eqref{eq:master}, the applied tidal field $P$ and the response $Q$, which at zero frequency grow as $r^{\ell+1}$ and decay as $r^{-\ell}$ respectively.
Any exterior solution is a combination of these two,
\begin{equation}\label{eq:split}
u=P+F\left(\omega\right)\,Q,
\end{equation}
where $F(\omega)$ is the \emph{response function}.
At low frequency, it has the expansion
\begin{equation}\label{eq:Fexp}
F(\omega)=\lambda_0+i\omega\lambda_1+\omega^2\lambda_2+i\omega^3\lambda_3+\omega^4\lambda_4+\dots \,, 
\end{equation}
whose coefficients $\lambda_n$ are the Love numbers.

\subsection{The surface impedance}

A body influences its exterior only through the boundary condition it imposes at its surface~\cite{Maggio:2020jml}.
At each frequency this boundary condition is one number: the ratio of the slope of the field to its value at the surface.
We define the \emph{surface impedance} as this ratio, normalized by the frequency,
\begin{equation}\label{eq:Z}
Z\equiv\frac{i}{\omega}\,\frac{u'}{u}\bigg|_{r=\Rb},
\end{equation}
where a prime denotes $\mathrm{d}/\mathrm{d}r_*$.
Three surfaces serve as reference points.
An infinitely stiff surface does not let the field move at all, $u(\Rb)=0$ (a Dirichlet condition), and has $Z=\infty$.
A free surface offers no resistance, $u'(\Rb)=0$ (a Neumann condition), and has $Z=0$.
A black hole horizon lets every incoming wave through and sends nothing back: the field there is purely ingoing, $u\propto e^{-i\omega(t+r_*)}$, so $u'=-i\omega u$ and $Z=1$ at the horizon, $\Rb=\Rs$.

It is convenient to define 
\begin{equation}\label{eq:H}
H\equiv i\omega Z=\left.-\frac{u^{\prime}}{u}\right|_{r=\Rb}\,.
\end{equation}
Since $Z$ and $H$ differ only by the factor of $i \omega$, we will refer to either as the impedance. 

Substituting the exterior solution \eqref{eq:split} into Eq.~\eqref{eq:H} gives the impedance in terms of the response function,
\begin{equation}\label{eq:ZofF}
H=-\frac{P'+F\,Q'}{P+F\,Q}\bigg|_{r=\Rb}\;.
\end{equation}
Solving for $F$ gives the response function in terms of the impedance,
\begin{equation}\label{eq:mobius}
F=-\frac{P'+H P}{Q'+H Q}\bigg|_{r=\Rb}\;.
\end{equation}
Each is a M\"obius map, a ratio of two linear functions, with coefficients that depend only on the exterior, and the two maps are inverse to each other.
The impedance and the response function therefore carry the same information.

\subsection{Two properties of the body}
\label{sec:propA}

Throughout, we consider stable bodies, for which no perturbation grows exponentially in time.
Without a model of the interior, the impedance is restricted only by general requirements, such as causality and stability.
Two physical properties, each shared by broad classes of astrophysical bodies, already restrict it strongly, and neither property refers to what the body is made of.

Both properties concern the energy that flows into the body, which we now relate to the impedance.
The field $u$ carries the energy
\begin{equation}\label{eq:energy}
E=\frac12\int_{\Rb}^\infty\big(\dot u^2+u'^2+Vu^2\big)\,\mathrm{d}r_*,
\end{equation}
which is positive since $V>0$ outside the horizon; far from the body it is the energy of gravitational waves.
Apart from radiation at infinity, $E$ changes only through the surface, $\mathrm{d}E/\mathrm{d}t=-p$, where $p=\dot u\,u'|_{r=\Rb}$ is the power that flows into the body.
Evaluating the power on the physical field, $\re[u(r)e^{-i\omega t}]$, and averaging over the phase of $u$ gives 
\begin{equation}\label{eq:power}
\langle p\rangle=\tfrac12|\omega|^2|u(\Rb)|^2e^{2t\im\omega}\,\im\frac{H(\omega)}{\omega}.
\end{equation}

\phantomsection\label{prop:a}\textbf{Lossless bodies.}\ The first property is that the body dissipates no energy; we call such a body lossless.
It holds for any body in a stable equilibrium whose internal friction is negligible on the time scale of the tide.
Neutron stars in an inspiral are an example: viscous dissipation is too weak to affect the tide~\cite{Bildsten:1992my,Kochanek:1992wk}.
Mathematically, we describe a lossless body by an impedance that is a sum, with positive weights, over the modes of the body with an infinitely stiff surface, $u(\Rb)=0$, taking the form 
\begin{equation}\label{eq:modesum}
H=-c+c_\infty\,\omega^2+\sum_{n\ge1} c_n\,\frac{\omega^2}{\Omega_n^2(\Omega_n^2-\omega^2)},\qquad c_n>0,\quad c_\infty\ge0,\quad\Omega_n^2>0\,.
\end{equation}
The poles are at the frequencies $\Omega_n$ of these modes, where $u(\Rb)=0$ and $H=-u'/u$ diverges. 
They are labeled in increasing order, $0<\Omega_1<\Omega_2<\cdots$, and stability ensures $\Omega_n^2>0$: a pole at $\Omega_n^2<0$ would give the body a mode that grows exponentially in time (Appendix~\ref{app:stability}).
The weight $c_n$ measures how strongly the $n$-th mode couples to the surface; positive weights mean that $H$ increases with $\omega^2$ between its poles.
A \emph{Herglotz} function $H(\omega^2)$ is a holomorphic map from the open upper half-plane $\im\omega^2>0$ to the closed upper half-plane $\im H(\omega^2)\geq0$~\cite{Donoghue:1974}.
In these terms, Eq.~\eqref{eq:modesum} says that $H$ is a Herglotz function of $\omega^2$ that is real on the real axis with poles only at $\omega^2>0$~\cite{Foster:1924,Wigner:1951,Luger:2022}.
We take Eq.~\eqref{eq:modesum} as the definition of a lossless body; it holds for any stable body whose interior conserves energy and has a discrete spectrum~\cite{Wigner:1951}.
The constant $c=-H(0)$, the stiffness of the surface against a static tide, can have either sign.

\phantomsection\label{prop:b}\textbf{Passive bodies.}\ The second property is that the body never gives back more energy than it has received; we call such a body passive.
A black hole is an example, since it absorbs energy but never emits any.
On the other hand, a body that releases energy of its own when the tide acts on it, such as an unstable body, is not passive.
Physically, we consider a perturbing field $\re[u(r)e^{-i\omega t}]$ that vanishes in the distant past.
This means that $\im\omega>0$.
Passivity then requires the energy received by the body from the tide up to any time $t$ to be non-negative:
\begin{equation}
\int_{-\infty}^t\langle p\rangle\,\mathrm{d}t'=\frac{|\omega|^2\,|u(\Rb)|^2\,e^{2t\im\omega}}{4\,\im\omega}\;\im\frac{H(\omega)}{\omega}\;\ge0 \,,
\end{equation}
where Eq.~\eqref{eq:power} was used.
A passive body therefore has 
\begin{equation}\label{eq:PR}
\im\frac{H(\omega)}{\omega}\ge0\qquad\text{for}\qquad\im\omega>0 \, .
\end{equation}
Near a pole $H$ takes values of every phase, so Eq.~\eqref{eq:PR} excludes poles of $H$ where $\im\omega>0$, making $H$ analytic there.
A real field produces a real response, so $H$ is real when $\omega$ is imaginary.
Together, these properties say that $Z=-iH/\omega$ is \emph{positive-real}~\cite{Brune:1931}, and the equivalence of positive-reality with passivity is shown in Refs.~\cite{YoulaCastriotaCarlin1959,Bernland:2011,Luger:2022}.
At low frequency, $H\simeq -c$ so Eq.~\eqref{eq:PR} requires the surface to have non-negative stiffness against a static tide,
\begin{equation}\label{eq:staticslope}
c\ge0 .
\end{equation}
We define passive bodies with respect to the energy \eqref{eq:energy}, with no surface term added at $r=\Rb$.
Appendix~\ref{app:energy} discusses this choice and what changes if such a term is added.

The two properties overlap.
A lossless body is passive exactly when $c\ge0$, since each term of Eq.~\eqref{eq:modesum}, divided by $\omega$, then has a non-negative imaginary part wherever $\im\omega>0$.
Conversely, a passive body that absorbs no energy at real frequencies is lossless~\cite{Foster:1924,Brune:1931,Luger:2022}.

These properties have direct analogues in positivity bounds: the body plays the role of the unknown ultraviolet completion, passivity that of unitarity and causality, and the mode sum of a lossless body that of a tree-level completion, a sum of narrow resonances with positive residues.

\subsection{The bodies we study}
\label{sec:bodies}

\begin{table}[t]\centering\small
\setlength{\tabcolsep}{4pt}
\begin{tabular}[t]{l c c}
\toprule
reference body & lossless & passive\\
\midrule
infinitely stiff surface & \checkmark & \checkmark\\
free surface & \checkmark & \checkmark\\
black hole & $\times$ & \checkmark\\
\bottomrule
\end{tabular}
\hspace{1.5em}
\begin{tabular}[t]{l c c}
\toprule
example body & lossless & passive\\
\midrule
SLy stars & \checkmark & --\\
polytropes & \checkmark & --\\
wormhole & -- & \checkmark*\\
thin-shell gravastar & -- & \checkmark*\\
\bottomrule
\end{tabular}
\caption{The reference bodies have constant impedances: the infinitely stiff surface ($Z=\infty$) and the free surface ($Z=0$) are the two extremes and reflect every wave, and the black hole ($Z=1$) absorbs every wave. We test our bounds on the example bodies. A body has the properties marked with \checkmark\ and not those marked with $\times$; a dash means that we do not assume the property for that body. For the wormhole and the zero-density thin-shell gravastar of Ref.~\cite{Cardoso:2017cfl}, the mark \checkmark*\ records only the static condition $c\ge0$ (Sec.~\ref{sec:bodies}), which is necessary but not sufficient for passivity. Indeed, with its incompressible shell, this gravastar has a mode that grows exponentially in time, so it is not passive.}
\label{tab:examples}
\end{table}

Table~\ref{tab:examples} lists the bodies we consider and the properties they have.
Both properties can be checked directly for the reference bodies.
The data for the stars we use are in Table~\ref{tab:stars} in Appendix~\ref{app:stars}.
These are built from the SLy equation of state~\cite{Douchin:2001sv,Read:2008iy} and from polytropes of index $n=0.5$, $1$ and $1.5$.
The polytropes have pressure $K\rho^{1+1/n}$, where $\rho$ is the rest-mass density and $K$ a constant.\footnote{For polytropes defined with the energy density, rather than the rest-mass density we use here, the values in Table~\ref{tab:stars} change, but our bounds hold for every star and the earlier bounds again fail for compact stars.}

The impedance of a perfect-fluid star under axial perturbations takes the form \eqref{eq:modesum}. 
This is because the interior obeys a wave equation~\cite{ChandrasekharFerrari1991c},
\begin{equation}\label{eq:interior}
    u^{\prime\prime}+(\omega^2-V_{\rm in})u=0\,,
\end{equation}
where the prime is a derivative with respect to the tortoise coordinate inside the body.
The potential $V_{\rm in}$ is real because a perfect fluid has no internal friction, which would add an imaginary part that absorbs energy.
Multiplying by the complex conjugate $\bar u$ and integrating from the center of the body to its surface gives
\begin{equation}\label{eq:herglotz}
\im H=\im(\omega^2)\,\frac{\int_{r_*(0)}^{r_*(\Rb)}|u|^2\,\mathrm{d}r_*}{|u(\Rb)|^2}\,,
\end{equation}
where there is no contribution from the center because the solution is regular there.
At real $\omega^2$ the right-hand side is zero, so $H$ is real and the average power \eqref{eq:power} vanishes: the star dissipates no energy.
Off the real axis the ratio is positive, so $\im H$ has the sign of $\im\omega^2$.
The star's interior is finite and the regular solution depends analytically on $\omega^2$, so $u(\Rb)$ and $u'(\Rb)$ are entire functions of $\omega^2$.
Consequently, $H=-u'/u$ is meromorphic in $\omega^2$, with poles only where $u(\Rb)=0$.
With stability, these properties give the mode sum~\eqref{eq:modesum} (Sec.~\ref{sec:propA}).

For polar tides, on the other hand, the fluid and the metric remain coupled inside the star, and there is no single wave equation of the form \eqref{eq:interior}.
Nevertheless, for each of the 22 stars in Table~\ref{tab:stars} we check numerically that the two poles of $H$ nearest $\omega^2=0$ in the complex plane lie at $\omega^2>0$ and have positive weights, as for a lossless body.
In Newtonian gravity every star is lossless~\cite{Chakrabarti:2013xza,Andersson:2019ahb}.

Polytropes can fail to be passive for polar perturbations, at compactness $C=M/\Rb$ below $0.016$ for $n=1.5$, below $0.101$ for $n=1$, and below $0.199$ for $n=0.5$.
This reflects our choice of the Zerilli function as the variable, not the star itself: since there is no single polar wave equation inside, the energy~\eqref{eq:energy}, built from the Zerilli function alone, differs from the physical energy of the perturbation by a surface term that can have either sign. 
Consequently, we do not assume that stars are passive.

For the wormholes and the zero-density thin-shell gravastars of Ref.~\cite{Cardoso:2017cfl}, we verify only the static condition~\eqref{eq:staticslope}, $c\geq0$.
This condition is necessary for passivity but not sufficient, so we mark it with \checkmark* in Table~\ref{tab:examples}.

\section{From the impedance to Love numbers}
\label{sec:schemes}

Our goal is to turn the properties of the impedance introduced in the previous section into bounds on Love numbers.
The impedance fixes the response function, and with it the Love numbers, through the map \eqref{eq:mobius}, which depends on how the exterior field is split into tide and response.
The simplest route is to split the field at the body's surface, in a scheme in which the response function inherits the mode sum of a lossless body directly.
We then translate the resulting bounds into bounds on the Love numbers of the worldline EFT, at any renormalization scale.

\subsection{Freedom in splitting tide and response}
\label{sec:splitting_freedom}

At fixed frequency, the exterior equation \eqref{eq:master} is a linear second-order ordinary differential equation. Its solutions form a two-dimensional space. 
The body contributes one condition, the impedance. This selects a one-dimensional subspace and the body's field $u$ is then unique up to its amplitude. 

Love numbers are defined by a further split of $u$, as in Eq.~\eqref{eq:split}. 
This requires a choice of two basis solutions, the tide $P$ and the response $Q$, and the response function $F$ is the coefficient of the response once the tide is normalized to one. 
So far, ``tide'' and ``response'' are just labels: any two linearly independent solutions to Eq.~\eqref{eq:master} are equally valid. 
Physically, large $r$ singles out a split: the tide grows and is the external field modified only by the body's mass, while the response decays and is generated by the body's finite size.
We impose this on $P$ and $Q$ and refer to it as the \emph{large-$r$ criterion}. 
In the worldline EFT the same distinction is between diagrams without and with an insertion of a finite-size operator, whose coefficients are the Love numbers.

In Newtonian gravity, the tide is exactly $r^{\ell+1}$ and the response $r^{-\ell}$ (in the variable $u$), and the two never mix. 
In general relativity, the tide scatters off the body's mass, generating a series of relativistic corrections, \begin{equation}
\label{eq:PN_corrections}
    P^{(0)}=r^{\ell+1}\sum_na_n(\Rs/r)^n \qquad \text{and} \qquad Q^{(0)}=r^{-\ell}\sum_nb_n(\Rs/r)^n\;,
\end{equation}
where $a_0=b_0=1$. 
The axial static solutions follow from the closed forms of Refs.~\cite{Binnington:2009bb,Damour:2009vw}, and the polar ones from these by the map between the Regge--Wheeler and Zerilli equations~\cite{Hui:2020xxx}.
For $\ell=2$, in units $\Rs=1$ and normalized so that $P^{(0)}\to r^3$ and $Q^{(0)}\to r^{-2}$, they are
\begin{equation}\label{eq:P0Q0}
\begin{aligned}
&\text{polar:}\quad P^{(0)}=\frac{r(4r^3+6r^2-3)}{4r+3},\\
&\phantom{\text{polar:}}\quad Q^{(0)}=-\frac{5r\left[12r^2+24r+13-3(4r^3+6r^2-3)\ln\frac{r}{r-1}\right]}{12r+9},\\
&\text{axial:}\quad P^{(0)}=r^3,\\
&\phantom{\text{axial:}}\quad Q^{(0)}=\frac{5}{12r}\left[12r^4\ln\frac{r}{r-1}-12r^3-6r^2-4r-3\right].
\end{aligned}
\end{equation}
The growing series, $P^{(0)}$, reaches the power of the decaying one, $Q^{(0)}$, at $n=2\ell+1$, mixing tide and response.
This ambiguity is harmless in the static case: for non-integer $\ell$ the two series share no power, and dimensional regularization makes $\ell$ effectively non-integer~\cite{Kol:2011vg}. 
The worldline EFT achieves the same result from the other side: the nonlinear corrections to the field of a point mass that reach the response power are finite and computable, and are subtracted in the matching. 
These two prescriptions agree and leave no free parameter, so at zero frequency the split is unique and $\lambda_0$ is unambiguous. 

We now impose the large-$r$ criterion at finite frequency, order by order, by expanding the basis functions in frequency 
\begin{equation}
    P=P^{(0)}+\omega^2P^{(2)}+\dots\;,\qquad Q=Q^{(0)}+\omega^2Q^{(2)}+\dots\;,
\end{equation}
and solving Eq.~\eqref{eq:master} perturbatively in $\omega$. Doing so generically produces logarithms: 
\begin{subequations}
\label{eq:log_ambiguity}
\begin{equation}
P=P^{(0)}+\omega^2\Big[\,r^{\ell+3}\sum_{n\ge0}a^{(2)}_n\Big(\frac{\Rs}{r}\Big)^n-\nu_2\Rs^2\,P^{(0)}\ln\frac{r}{r_0}+\sigma \Rs^{2\ell+3}\,Q^{(0)}\ln\frac{r}{r_0}\Big]+O(\omega^4),
\end{equation}
\begin{equation}
Q=Q^{(0)}+\omega^2\Big[\,r^{2-\ell}\sum_{n\ge0}b^{(2)}_n\Big(\frac{\Rs}{r}\Big)^n+\nu_2\Rs^2\,Q^{(0)}\ln\frac{r}{r_0}\Big]+O(\omega^4)\;,
\end{equation}
\end{subequations}
where $\nu_2$ and $\sigma$ are pure numbers fixed by the exterior geometry.\footnote{$-\nu_2$ is the coefficient of $(\Rs\omega)^2$ in the renormalized angular momentum $\nu-\ell$ of Ref.~\cite{Mano:1996vt}, and $\sigma\Rs^{2\ell+2}$ is the leading absorptive Love number $\lambda_1$ of a black hole~\cite{Combaluzier--Szteinsznaider:2025eoc}. For $\ell=2$, $\nu_2=107/210$ and $\sigma=1/5$ in both parities.} 

The logarithms introduce a reference length $r_0$, which is an artifact of the split in Eq.~\eqref{eq:split} and drops out of the full $u$. 
Changing $r_0$ mixes tide and response and leads to different basis functions.
In the worldline EFT the logarithms accompany divergences: the worldline's finite-size operators act as counterterms, their finite parts are a choice, and the inverse renormalization scale $1/\mu$ plays the role of the reference length $r_0$~\cite{Caron-Huot:2025tlq}.

At each order in $\omega^2$ the corrections are fixed only up to adding the static solutions $P^{(0)}$ and $Q^{(0)}$, which solve the equation without the source.
The large-$r$ criterion forbids adding $P^{(0)}$ to the response, which would then grow, but leaves three coefficients in Eq.~\eqref{eq:log_ambiguity} undetermined: $a^{(2)}_2$ and $a^{(2)}_{2\ell+3}$, at the powers of $P^{(0)}$ and $Q^{(0)}$ in $P^{(2)}$, and $b^{(2)}_2$, at the power of $Q^{(0)}$ in $Q^{(2)}$, which are also where the logarithms appear.
With $r_0=1/\mu$, we fix them as in the worldline EFT in dimensional regularization with the subtraction convention of Ref.~\cite{Caron-Huot:2025tlq}, used in Refs.~\cite{Combaluzier--Szteinsznaider:2025eoc,Apostolidis:2026qsg}, so that our $\lambda_2$ is the Wilson coefficient of those references. This leaves $r_0$ as the only freedom.
We call this choice of basis, set by the large-$r$ criterion and labeled by $r_0$, the \emph{asymptotic scheme}.

The consequence for the response is that dynamical Love numbers run for $n\geq2$. 
Since $u$ does not depend on $r_0$, substituting Eq.~\eqref{eq:log_ambiguity} into Eq.~\eqref{eq:split} relates the Love numbers at two scales $r_0$ and $r_0'$~\cite{Saketh:2023bul,Combaluzier--Szteinsznaider:2025eoc,Kosmopoulos:2025rfj}:
\begin{equation}
\lambda_0(r_0')=\lambda_0(r_0),\qquad \lambda_1(r_0')=\lambda_1(r_0),\qquad
\lambda_2(r_0')=\lambda_2(r_0)+\big(\sigma \Rs^{2\ell+1}+2\nu_2\lambda_0\big)\Rs^2\ln\frac{r_0'}{r_0}\;,
\label{eq:running}
\end{equation}
where the higher dynamical Love numbers generally also depend on $\ln(r_0'/r_0)$.

\subsection{The surface scheme}
\label{sec:static_scheme}

The response function inherits the mode sum \eqref{eq:modesum} from $H$ if the map \eqref{eq:mobius} has coefficients that are real and independent of frequency.
The asymptotic scheme does not provide such a map, since the values and slopes of its basis functions at the surface depend on $\omega$.
We therefore relax the large-$r$ criterion and consider a broader notion of scheme.

A solution of Eq.~\eqref{eq:master} is fixed by its value and slope at one radius, which we take to be the body's surface $r=\Rb$, where the impedance is defined.
Four functions of frequency, $P(\Rb)$, $P'(\Rb)$, $Q(\Rb)$ and $Q'(\Rb)$, thus determine a pair of basis functions at every $r$.
We require the two basis functions $P$ and $Q$ to satisfy three conditions:
\begin{enumerate}\label{list:basis_conditions}
    \item They reduce to the unique static pair \eqref{eq:PN_corrections} as $\omega\to0$.
    \item They are linearly independent.
    \item They are real and analytic in $\omega^2$, so that the basis adds no odd powers of $\omega$ to $F$ beyond those of $H$.
\end{enumerate}
We call such a choice of the four functions a \emph{scheme}; any scheme leaves $u$ unchanged and changes only how it is divided into tide and response.
The asymptotic scheme of Sec.~\ref{sec:splitting_freedom}, fixed by the large-$r$ criterion, carries the scale $r_0$ because that criterion is a condition at large $r$ rather than data at $\Rb$.
Any other choice of the four functions that meets the three conditions is also a scheme, so the notion is broader than a renormalization scheme in the EFT sense.

We call the scheme that carries the mode sum to the response function the \emph{surface scheme}.
Its basis functions $\tilde P$ and $\tilde Q$ have the values and slopes of the static solutions at the body's surface,
\begin{subequations}
\begin{equation}
\tilde P(\Rb)=P^{(0)}(\Rb),\qquad \tilde P'(\Rb)=P^{(0)\prime}(\Rb),
\end{equation}
\begin{equation}
\tilde Q(\Rb)=Q^{(0)}(\Rb),\qquad \tilde Q'(\Rb)=Q^{(0)\prime}(\Rb).
\end{equation}
\label{eq:staticdata}
\end{subequations}
They are exact solutions at every frequency and satisfy the three conditions.
The body's field is then
\begin{equation}
\label{eq:static_split}
    u=A(\omega)\,\big(\tilde P+g\,\tilde Q\big)
\end{equation}
at every $r>\Rb$, where the normalization $A(\omega)$ differs from that of Eq.~\eqref{eq:split} because the tide is normalized differently.
By Eq.~\eqref{eq:mobius}, the response function in the surface scheme is
\begin{equation}
\label{eq:gmap}
g(\omega)=-\,\frac{P^{(0)\prime}+H(\omega)\,P^{(0)}}{Q^{(0)\prime}+H(\omega)\,Q^{(0)}}\bigg|_{r=\Rb}\;.
\end{equation}
This is a M\"obius map in $H$ whose coefficients, the values and slopes of $P^{(0)}$ and $Q^{(0)}$ at $\Rb$, are real and independent of frequency.
No other scheme has this property: by condition~1, frequency-independent values at $\Rb$ must be the static ones, up to a common factor that rescales both basis functions and leaves $g$ unchanged.

For a lossless body, $H$ is a Herglotz function of $\omega^2$
(Sec.~\ref{sec:propA}). 
Through the map~\eqref{eq:gmap}, $g(\omega^2)$ inherits the Herglotz property of $H(\omega^2)$ because 
\begin{equation}
    \im g=\frac{2\ell+1}{\big|Q^{(0)\prime}+HQ^{(0)}\big|^2}\,\im H\,,
\end{equation}
where the numerator is minus the Wronskian $W[f,h]\equiv fh'-f'h$ of the static pair, $W[P^{(0)},Q^{(0)}]=-(2\ell+1)$.
Therefore, $g$ has non-negative imaginary part wherever $H$ does, and it is holomorphic there because the denominator vanishes only where $H$ is real.

\subsection{Changing schemes}
\label{sec:change}

Love numbers are usually quoted in the asymptotic scheme, for example as the Wilson coefficients of the worldline EFT.
On the other hand, the surface scheme is fixed at the body's surface rather than at large $r$.
We therefore translate bounds derived in the surface scheme to the asymptotic scheme, at any scale $r_0$.

In the asymptotic scheme the basis is $(P,Q)$ and the response function is $F$.
At $r=\Rb$ the change of basis reads
\begin{subequations}
\label{eq:decomp}
\begin{equation}
\begin{pmatrix}P(\Rb)\\ P'(\Rb)\end{pmatrix}
=m_{11}\begin{pmatrix}P^{(0)}(\Rb)\\ P^{(0)\prime}(\Rb)\end{pmatrix}
+m_{21}\begin{pmatrix}Q^{(0)}(\Rb)\\ Q^{(0)\prime}(\Rb)\end{pmatrix},
\end{equation}
\begin{equation}
\begin{pmatrix}Q(\Rb)\\ Q'(\Rb)\end{pmatrix}
=m_{12}\begin{pmatrix}P^{(0)}(\Rb)\\ P^{(0)\prime}(\Rb)\end{pmatrix}
+m_{22}\begin{pmatrix}Q^{(0)}(\Rb)\\ Q^{(0)\prime}(\Rb)\end{pmatrix},
\end{equation}
\end{subequations}
with coefficients given by Wronskians at $\Rb$,
\begin{equation}
\begin{aligned}
m_{11}&=\frac{W[P,Q^{(0)}]}{W[P^{(0)},Q^{(0)}]},&\qquad m_{12}&=\frac{W[Q,Q^{(0)}]}{W[P^{(0)},Q^{(0)}]},\\
m_{21}&=\frac{W[P^{(0)},P]}{W[P^{(0)},Q^{(0)}]},&\qquad m_{22}&=\frac{W[P^{(0)},Q]}{W[P^{(0)},Q^{(0)}]}.
\end{aligned}
\label{eq:coeffs}
\end{equation}
Adding the first line of Eq.~\eqref{eq:decomp} to $F$ times the second, and using that a solution is fixed by its value and slope at $\Rb$, gives
\begin{equation}
    u=P+FQ=(m_{11}+m_{12}F)\,\tilde P+(m_{21}+m_{22}F)\,\tilde Q
\end{equation}
at every $r>\Rb$.
Comparing with Eq.~\eqref{eq:static_split} gives the exact map
\begin{equation}
g=\frac{m_{21}+m_{22}F}{m_{11}+m_{12}F}\;,\qquad\qquad F=\frac{m_{11}\,g-m_{21}}{m_{22}-m_{12}\,g}\;.
\label{eq:transport}
\end{equation}
Its coefficients come from the exterior alone: they depend on $\omega$, on $\Rb$ and on $r_0$, but not on the interior, so the same map converts the response function of every body with that radius.
Because they depend on $\omega$, the map does not in general preserve the Herglotz property of $g$.

The consequence for the Love numbers follows by expanding in $\omega$.
At leading order, by conditions~1 and~3 of Sec.~\ref{sec:static_scheme}, $m_{11}=1+a_{11}\Rs^2\omega^2$, $m_{12}=a_{12}\Rs^2\omega^2$, $m_{21}=a_{21}\Rs^2\omega^2$ and $m_{22}=1+a_{22}\Rs^2\omega^2$, up to $O(\omega^4)$.
Writing the small frequency expansion 
\begin{equation}\label{eq:g_expansion}
    g=g_0+i\omega g_1+\omega^2g_2+\dots\,,
\end{equation}
the map \eqref{eq:transport} gives
\begin{equation}
\lambda_0=g_0,\qquad \lambda_1=g_1,\qquad \lambda_2=g_2-\Rs^2\big[a_{21}+(a_{22}-a_{11})\lambda_0-a_{12}\lambda_0^2\big]\;.
\label{eq:g2shift}
\end{equation}
The static Love number and $\lambda_1$ are thus the same in every scheme and at every scale, while the Love numbers with $n\ge2$ generally depend on both.
For a lossless body, $H$ has no odd powers of $\omega$, so $\lambda_1=g_1=0$ and every odd coefficient vanishes in every scheme.

Equation~\eqref{eq:g2shift} splits $\lambda_2$ into two parts.
The first, $g_2$, is the body's own dynamical response: it vanishes when the impedance does not depend on frequency, $H(\omega)=H(0)$, since $g$ is then constant by Eq.~\eqref{eq:gmap}.
The second involves the body only through $\lambda_0$.
It is the dynamical Love number of such a frozen body, whose interior does not respond at finite frequency,
\begin{equation}
\lambda_2^{\rm frozen}(\lambda_0)=-\Rs^2\big[a_{21}+(a_{22}-a_{11})\lambda_0-a_{12}\lambda_0^2\big]\;,
\label{eq:robin}
\end{equation}
and it comes from the propagation of the field between the surface and infinity.
This is a relativistic effect; it is absent in Newtonian gravity, where the exterior solutions do not depend on frequency.
Thus
\begin{equation}
\lambda_2=g_2+\lambda_2^{\rm frozen}(\lambda_0)\;,
\label{eq:g2inv}
\end{equation}
with $\lambda_2^{\rm frozen}=0$ in the surface scheme.
The frozen value carries all the dependence on the scheme and the scale, and it is the same for every body with the same radius and static Love number.
The body's contribution, $g_2$, depends on neither and is what we bound for lossless bodies in Sec.~\ref{sec:bounds}.
Expanding Eq.~\eqref{eq:transport} to higher orders in $\omega$ relates each higher Love number to the corresponding coefficient of $g$ in the same way.

Explicitly, by first-order perturbation theory in $\omega^2$,
\begin{equation}\label{eq:frozenint}
\lambda_2^{\rm frozen}(\lambda_0)=\frac{1}{2\ell+1}\,{\rm FP}\!\int_{\Rb}^{\infty}\big[P^{(0)}+\lambda_0Q^{(0)}\big]^2\,\mathrm{d}r_*\;+\;q_0\Rs^{2\ell+3}+q_1\lambda_0\Rs^2 .
\end{equation}
The integral diverges at large $r$, and FP keeps its finite part: with a cutoff at $r=\Lambda$, it discards the positive powers of $\Lambda$ and the logarithm $(2\ell+1)\big(\sigma\Rs^{2\ell+1}+2\nu_2\lambda_0\big)\Rs^2\ln(\Lambda/r_0)$, which gives the running \eqref{eq:running}.
The constants $q_0$ and $q_1$ fix the convention of Sec.~\ref{sec:splitting_freedom}.
For $\ell=2$, $q_0=5715539/25804800$ and $q_1=10427/88200$ for polar tides, and $q_0=2071/12600$ and $q_1=323/22050$ for axial tides.

To derive Eq.~\eqref{eq:frozenint}, write $u_0=P^{(0)}+\lambda_0Q^{(0)}$ for the frozen body's static field and $u_2=P^{(2)}+\lambda_0Q^{(2)}$.
The coefficients \eqref{eq:coeffs} give $\lambda_2^{\rm frozen}=W[u_0,u_2]/(2\ell+1)$ at $\Rb$, and Eq.~\eqref{eq:master} at order $\omega^2$ gives $\mathrm{d}W[u_0,u_2]/\mathrm{d}r_*=-u_0^2$.
Integrating from $\Rb$ to infinity gives the integral in Eq.~\eqref{eq:frozenint}, and the boundary term at infinity gives $q_0$ and $q_1$.

\section{Bounds on Love numbers}
\label{sec:bounds}

We now derive bounds on Love numbers for bodies that are passive, lossless, or both, first in the surface scheme and then in any scheme through Eq.~\eqref{eq:g2inv}.

\subsection{The static Love number}
\label{sec:window}

The static Love number of a passive body or a stable lossless body is bounded from below by that of an infinitely stiff surface of the same radius.
For a passive body, the bound becomes two-sided, and its ceiling is the static Love number of a free surface.
A passive body has $c=-H(0)\ge0$ by Eq.~\eqref{eq:staticslope}.
This carries over to the static Love number through the map~\eqref{eq:gmap} evaluated at $\omega=0$:\footnote{For a massless scalar field rather than gravitational perturbations, Ref.~\cite{Giddings:2026oix} writes the static Love number in the same way, in terms of the static response at a surface.}
\begin{equation}\label{eq:lam0ofc}
\lambda_0=-\frac{P^{(0)\prime}-c\,P^{(0)}}{Q^{(0)\prime}-c\,Q^{(0)}}\bigg|_{r=\Rb}\,.
\end{equation}
The map \eqref{eq:lam0ofc} is monotone on any interval on which its denominator does not vanish, and $c\ge0$ is such an interval: the pole at $c=Q^{(0)\prime}/Q^{(0)}|_{\Rb}$ is negative because $Q^{(0)}$ is positive, decays at large $r$, and is convex where $V>0$.
Consequently, as $c$ increases from $0$ to $\infty$, $\lambda_0$ decreases (Sec.~\ref{sec:static_scheme}) from the Love number of a free surface, $Z=0$, to that of an infinitely stiff surface, $Z=\infty$ (see Fig.~\ref{fig:Hg}, right).
Therefore, a passive body has
\begin{equation}\label{eq:window}
\lamStiff(\Rb)\le\lambda_0\le\lamFree(\Rb),\qquad \lamStiff=-\frac{P^{(0)}}{Q^{(0)}}\bigg|_{\Rb},\qquad \lamFree=-\frac{P^{(0)\prime}}{Q^{(0)\prime}}\bigg|_{\Rb},
\end{equation}
where $P^{(0)}$ and $Q^{(0)}$ are the static solutions \eqref{eq:P0Q0}, so the bounds are explicit functions of the compactness.
The lower bound holds for every passive body and every stable lossless body: a lossless body with $\lambda_0<\lamStiff$ has a mode that grows exponentially in time (Appendix~\ref{app:stability}).
The upper bound rests on the choice of energy at the surface (Appendix~\ref{app:energy}), while the lower bound does not.
The window holds for every multipole and both parities, with the corresponding static solutions.

For $\ell=2$ polar tides we quote the Love number normalized by the mass, $k^E_2=\lambda_0/(12M^5)$, as in Ref.~\cite{Cardoso:2017cfl}.
The Love number $k_2=\lambda_0/(12\Rb^5)$ of Refs.~\cite{Hinderer:2007mb,Binnington:2009bb} is $k_2=C^5k^E_2$ in our notation.
The window \eqref{eq:window} then reads 
\begin{equation}
    k_2^{E,\rm stiff}\le k^E_2\le k_2^{E,\rm free}\,,
\end{equation} 
with
{\medmuskip=2mu
\begin{align}
k_2^{E,\rm stiff}&=\frac{8\,(1+3C-6C^3)}{5\big[2C(3+12C+13C^2)+3(1+3C-6C^3)\ln(1-2C)\big]},\label{eq:edges}\\
k_2^{E,\rm free}&=\frac{8\,(1-2C)(2+8C+9C^2-6C^4)}{5\big[2C(6+18C-C^2-47C^3-44C^4)+3(1-2C)(2+8C+9C^2-6C^4)\ln(1-2C)\big]}.\nonumber
\end{align}}%
In the Newtonian limit, $C\to0$, the corresponding values of $k_2$ tend to $-1/12$ and $1/8$.
The lower bound is the Love number of the perfect mirror of Ref.~\cite{Cardoso:2017cfl}, and reproduces its values.

For a horizon-scale body the surface lies at $\Rb=\Rs(1+\delta)$ with $\delta\ll1$, so that $C=1/[2(1+\delta)]$.
As $\delta\to0$ the bounds \eqref{eq:edges} become
\begin{equation}\label{eq:windowhorizon}
k_2^{E,\rm stiff}=\frac{8}{5(7+3\ln\delta)},\qquad k_2^{E,\rm free}=\frac{72}{35}\,\delta,
\end{equation}
which are shown in Fig.~\ref{fig:love} (right).\footnote{The hatched regions in Fig.~\ref{fig:love} (right) are the 90\% bounds from GW250114~\cite{Andres-Carcasona:2025bni} on its primary, with a log-uniform prior, and from GW170608~\cite{Narikawa:2021pak}. For GW170608 we convert the bound on the binary's combination $\tilde\Lambda$ into a bound on each body by assuming equal masses and equal Love numbers.}
Both $k_2^{E,\rm stiff}$ and $k_2^{E,\rm free}$ vanish as $\delta\to0$, so a passive body whose surface lies at the horizon has the vanishing static Love number of the black hole.
The upper bound vanishes linearly in $\delta$ and the lower bound only logarithmically.
The wormhole and the thin-shell gravastar of Ref.~\cite{Cardoso:2017cfl} lie inside the window at every height.

We also tested the window on thin-shell gravastars whose shells are compressible, from Refs.~\cite{Pani:2009ss,Uchikata:2016qku}, including shells with nonzero surface energy density.
For each we computed the static Love number and searched for $\ell=2$ polar modes that grow exponentially in time.
Every gravastar we tested that lies below $\lamStiff$ has such a mode, consistent with the lower bound.
Those that lie above $\lamFree$ have no growing mode, but they have $c<0$ and so are not passive.
Some gravastars inside the window are also unstable so the lower bound is necessary for stability but not sufficient.

\subsection{The dynamical Love numbers}
\label{sec:dynamical}
\begin{figure}[t]
\centering
\includegraphics[width=\textwidth]{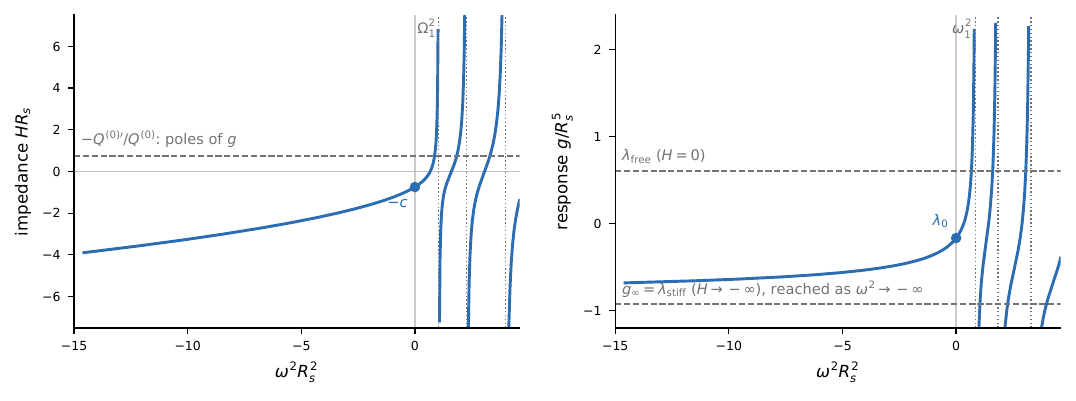}
\caption{LEFT: The impedance $H$ of the uniform-density star of Fig.~\ref{fig:qnm} (left), for axial tides with $\ell=2$, on the real $\omega^2$ axis.
It equals $-c$ at $\omega^2=0$ and increases between its poles (vertical dotted), which are the modes of the star with an infinitely stiff surface.
RIGHT: The response $g$ of the same star. It has a pole (vertical dotted) wherever $H$ crosses $-Q^{(0)\prime}/Q^{(0)}|_{\Rb}$ (dashed, left), increases between its poles, equals $\lambda_0$ at $\omega^2=0$, and approaches $\ginf=\lamStiff$ (dashed) as $\omega^2\to-\infty$. Its values at $H=0$ and $H\to-\infty$ are the edges $\lamFree$ and $\lamStiff$ of the window~\eqref{eq:window}.
The polar tides of the stars in Fig.~\ref{fig:love} show the same features, but with $\lambda_0>0$.}
\label{fig:Hg}
\end{figure}

For a lossless body, every dynamical Love number is bounded from above and below, given the static Love number, the mass and radius, and the lowest pole of the response.
The bounds follow because $g$ is also a Herglotz function of $\omega^2$ (Sec.~\ref{sec:static_scheme}), and takes the form
\begin{equation}\label{eq:gsum}
g(\omega)=\ginf+\sum_{k\ge1}\frac{\rho_k}{\omega_k^2-\omega^2},\qquad \rho_k>0,\quad 0<\omega_1<\omega_2<\cdots,
\end{equation}
where $\ginf\equiv g(\omega^2\to-\infty)$ is the response to arbitrarily fast tides and $\rho_k$ measures how strongly the $k$-th mode couples to the tide~\cite{Foster:1924,Wigner:1951,Luger:2022}.
Figure~\ref{fig:Hg} shows $H$ and $g$ for the uniform-density star of Fig.~\ref{fig:qnm}.
The poles of $g$ lie where $H$ crosses $-Q^{(0)\prime}/Q^{(0)}|_{\Rb}$, at which the denominator of Eq.~\eqref{eq:gmap} vanishes.
For $\omega^2\le0$, $H$ increases with $\omega^2$, so $-\infty<H\le-c$, and $-c<-Q^{(0)\prime}/Q^{(0)}|_{\Rb}$ because a stable lossless body has $\lambda_0>\lamStiff$ (Sec.~\ref{sec:window}).
Hence all poles of $g$ lie at $\omega^2>0$. For $\omega^2\le0$ the response $g$ stays between $\lamStiff$ and $\lambda_0$, so, unlike Eq.~\eqref{eq:modesum}, it has no term linear in $\omega^2$ and its limit $\ginf$ is finite with $\ginf\ge\lamStiff$.

In general relativity $\ginf$ takes the lowest value allowed by $\ginf\ge\lamStiff$, because the field crosses the body at finite speed. 
For a tide that grows on a time scale shorter than this crossing time, the wave returning from the interior is exponentially suppressed relative to the incoming one.
The field at the surface is then an ingoing wave, for which $Z=1$ and $H=i\omega Z\to-\infty$, so
\begin{equation}\label{eq:ginf}
\ginf=\lamStiff(\Rb),
\end{equation}
the Love number of an infinitely stiff surface.
We confirm Eq.~\eqref{eq:ginf} numerically for the polar tides of the stars of Fig.~\ref{fig:love}: it holds to $2\times10^{-8}$.
The value $\ginf$ is negative for every radius, since $P^{(0)}$ and $Q^{(0)}$ are positive outside the horizon.
It is also not small: for these stars $\ginf/\Rb^5$ lies between $-1.03$ and $-0.70$, while $\lambda_0/\Rb^5$ lies between $0.21$ and $3.9$.

Expanding Eq.~\eqref{eq:gsum} in $\omega^2$ and using Eq.~\eqref{eq:g_expansion} gives 
\begin{equation}
        \lambda_0-\ginf=\sum_k\frac{\rho_k}{\omega_k^2}\,,\qquad\qquad g_{2n}=\sum_k\frac{\rho_k}{\omega_k^{2n+2}}\,,\quad n\ge1\,.
\end{equation}
Two-sided bounds on $g_2$ then follow.
Positivity of each term in $g_2=\sum_k\rho_k/\omega_k^4$ gives the floor, and since every $\omega_k\ge\omega_1$, each term is at most $(\rho_k/\omega_k^2)/\omega_1^2$, which gives the ceiling.
Therefore, 
\begin{equation}\label{eq:chain}
    0\le\omega_1^2\,g_2\le\lambda_0-\ginf ,
\end{equation}
with equality on the right only when $g$ has a single pole. 
In any scheme, by Eq.~\eqref{eq:g2inv},
\begin{equation}\label{eq:chainlam}
\lambda_2^{\rm frozen}(\lambda_0)\;\le\;\lambda_2\;\le\;\lambda_2^{\rm frozen}(\lambda_0)+\frac{\lambda_0-\ginf}{\omega_1^2}.
\end{equation}
The same sums give further bounds, as in Ref.~\cite{Apostolidis:2026mvb}.
Each higher coefficient is bounded by the one below it, $0\le\omega_1^2\,g_{2n+2}\le g_{2n}$ for $n\ge1$.
The Cauchy--Schwarz inequality gives bounds of a different kind, for example $g_2^2\le(\lambda_0-\ginf)\,g_4$.

Figure~\ref{fig:love} (left) tests Eq.~\eqref{eq:chainlam} on the 22 stars in Table~\ref{tab:stars}, for polar tides with $\ell=2$, in the asymptotic scheme at $\mu=\re\omega_f$, where $\omega_f$ is the frequency of the star's fundamental mode, or \emph{$f$-mode}, its polar fluid mode of lowest frequency.
Both inequalities hold for every star.

For a Newtonian star $\lambda_2^{\rm frozen}=0$ (Sec.~\ref{sec:change}), the poles in Eq.~\eqref{eq:gsum} are the star's normal modes~\cite{Chakrabarti:2013xza,Andersson:2019ahb}, and $\ginf=0$.
The bound \eqref{eq:chain} then reads $0\le\omega_1^2\lambda_2\le\lambda_0$, which is the form of the earlier bounds~\cite{Kehagias:2026nqi,Apostolidis:2026mvb}.
Reference~\cite{Kehagias:2026opd} writes the ceiling as $\lambda_0-\lambda_\infty$, the same form as Eq.~\eqref{eq:chain} with $\lambda_\infty\ge0$ in place of $\ginf$.

The Newtonian value $\ginf=0$ differs from the limit $-\Rb^5$ of Eq.~\eqref{eq:ginf} as $C\to0$ because the limits $C\to0$ and $\omega^2\to-\infty$ do not commute.
The value $\ginf$ describes tides faster than the time the field takes to cross the body, and Newtonian gravity sets this time to zero.
At small $C$ the general-relativistic response therefore agrees with the Newtonian one up to corrections of order $C$ for tides slower than the crossing time and falls to $\lamStiff$ for faster tides.
A fast Newtonian tide finds the matter unable to respond.
Because the potential adjusts instantly, the field inside and outside the body is the applied tide alone.
The body is then transparent, $H=-P^{(0)\prime}/P^{(0)}|_{\Rb}$, and Eq.~\eqref{eq:gmap} gives $g=0$.
In general relativity, the field instead enters as an ingoing wave and $H\to-\infty$.

\begin{figure}[t]
\centering
\includegraphics[width=\textwidth]{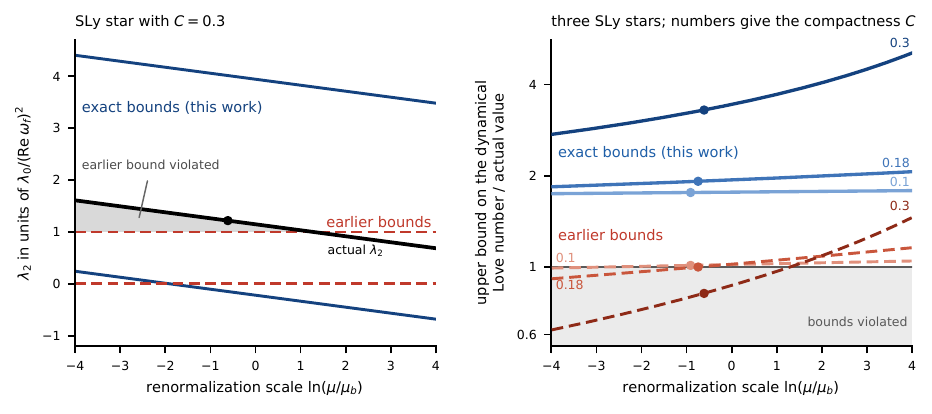}
\caption{LEFT: The leading dynamical Love number $\lambda_2$ of the SLy star with $C=0.3$ (black), in units of $\lambda_0/(\re\omega_f)^2$, versus $\ln(\mu/\mu_b)$, where $\mu=1/r_0$ is the renormalization scale of the asymptotic scheme and $\mu_b=1/\Rb$, together with our bounds \eqref{eq:chainlam} (blue) and the earlier bounds~\cite{Kehagias:2026nqi,Apostolidis:2026mvb} (red).
The earlier bounds are constant while $\lambda_2$ runs, so they fail at low scales (gray); ours move with $\lambda_2$.
RIGHT: The ratio of Fig.~\ref{fig:love} (left), upper bounds divided by the actual $\lambda_2$, versus the renormalization scale, for three SLy stars, including the star of $1.4\,M_\odot$ ($C=0.18$).
Dots mark $\mu=\re\omega_f$, the scale of Fig.~\ref{fig:love}.}
\label{fig:running}
\end{figure}

The earlier bounds differ from Eq.~\eqref{eq:chainlam} in two ways.
First, they take the response to arbitrarily fast tides to vanish~\cite{Kehagias:2026nqi,Apostolidis:2026mvb} or to be non-negative~\cite{Kehagias:2026opd}, whereas $\ginf<0$, which raises our upper bound.
Second, they are applied to $\lambda_2$ itself, at a fixed scale, with no frozen value subtracted~\cite{Kehagias:2026nqi,Apostolidis:2026mvb}.
In Newtonian gravity this costs nothing.
In general relativity $\lambda_2^{\rm frozen}$ carries the running of $\lambda_2$, so an inequality on $\lambda_2$ alone holds at some scales and fails at others.
The bound \eqref{eq:chainlam} is on $g_2=\lambda_2-\lambda_2^{\rm frozen}(\lambda_0)$, which does not depend on the scale, so it holds at every scale (Fig.~\ref{fig:running}).
The earlier bound $\lambda_0/(\re\omega_f)^2$, as applied in Refs.~\cite{Kehagias:2026nqi,Apostolidis:2026mvb}, fails at the scale tested in Fig.~\ref{fig:love}, $\mu=\re\omega_f$, for the stars with $C\ge0.2$, where $\lambda_2$ exceeds it by up to $22\%$.

Nonetheless, at low compactness the earlier bound lies close to $\lambda_2$, because of two properties of stars.
First, the fundamental mode accounts for most of the static Love number, $\rho_1/\omega_1^2\approx\lambda_0$, so by Eq.~\eqref{eq:gsum} the other poles together contribute about $-\ginf$.
Second, the other poles lie far above $\omega_1$ and change $g_2$ by less than $10\%$ for the 22 stars of Table~\ref{tab:stars}. 
Neither follows from the mode sum, and the earlier bound depends on both.
A counterexample is a surface with no modes, $H=-c+c_\infty\omega^2$ in Eq.~\eqref{eq:modesum}: its response $g$ has a single pole, at $\omega_1$, which carries all of $\lambda_0-\ginf$, so it saturates Eq.~\eqref{eq:chain} and exceeds the earlier bound.

The bounds \eqref{eq:chain} require the body to be lossless.
A body that absorbs, such as a black hole with $Z=1$ at its horizon~\cite{Damour:1978cg,Znajek:1978nwp,Price:1986yy,Thorne:1986iy}, has its response spread continuously down to zero frequency, so $g$ has no lowest pole and Eq.~\eqref{eq:chain} gives no upper bound~\cite{Kehagias:2026opd,Apostolidis:2026mvb}.

\subsection{Bodies with both properties}
\label{sec:both}

For a body with both properties, the upper bound $\lambda_0-\ginf$ in Eq.~\eqref{eq:chain} is at most $\lambda_0-\lamStiff(\Rb)$, since $\ginf\ge\lamStiff$.
Moreover, $\lambda_0\le\lamFree(\Rb)$ by the window \eqref{eq:window}, so
\begin{equation}\label{eq:combined}
0\le\omega_1^2\,g_2\le\lamFree(\Rb)-\lamStiff(\Rb),
\end{equation}
which does not involve the static Love number.
Of the stars in Fig.~\ref{fig:love}, 17 have $c\ge0$, and hence both properties, and satisfy Eq.~\eqref{eq:combined}, while 2 of the 5 stars with $c<0$ violate it.

\section{Bounds on quasinormal modes}
\label{sec:qnm}

A QNM is a frequency at which a body oscillates while radiating to infinity, with nothing coming in.
The radiated energy is lost, so the oscillation decays and the frequency is complex.
We show that, for lossless bodies, the QNMs are excluded from a region of the complex frequency plane that depends only on the mass and radius of the body.
Knowledge of the Love numbers enlarges this region.
Conversely, for such bodies, the QNMs bound each Love number from below, given the lower ones.
In particular, the mass, radius, static Love number and complex frequency of the fundamental mode bound the leading dynamical Love number from below.

\subsection{QNMs as a condition on the impedance}
\label{sec:QNM_impedance}

Let $u_{\rm out}$ be the exterior solution that is purely outgoing at infinity, and define the impedance that the exterior presents to the surface,
\begin{equation}\label{eq:Zout}
H_{\rm out}(\omega)\equiv-\frac{u_{\rm out}'}{u_{\rm out}}\bigg|_{\Rb}.
\end{equation}
The body's solution continues outward as a purely outgoing wave exactly when the two impedances agree, so the QNMs are the frequencies at which
\begin{equation}\label{eq:qnmcond}
H(\omega)=H_{\rm out}(\omega).
\end{equation}

The QNMs are also the poles of the partial-wave amplitude.
Write the exterior solution as
\begin{equation}
    u=A_{\rm in}u_{\rm in}+A_{\rm out}u_{\rm out}\;,
\end{equation}
where $u_{\rm in}\to e^{-i\omega r_*}$ and $u_{\rm out}\to e^{i\omega r_*}$ far from the body, and let $H_{\rm in}\equiv-u_{\rm in}'/u_{\rm in}|_{\Rb}$.
The partial-wave amplitude is
\begin{equation}\label{eq:Sell}
S_\ell=\frac{A_{\rm out}}{A_{\rm in}}=-\frac{u_{\rm in}}{u_{\rm out}}\bigg|_{\Rb}\,\frac{H-H_{\rm in}}{H-H_{\rm out}},
\end{equation}
a M\"obius image of $H$ with coefficients fixed by the exterior, which has a pole wherever Eq.~\eqref{eq:qnmcond} holds.

On the positive imaginary axis, the condition~\eqref{eq:qnmcond} describes modes that grow exponentially in time, and Appendix~\ref{app:stability} uses it to prove the lower bound of the window.

\subsection{Lossless bodies of given radius}
\label{sec:QNM_prop_a}

For a lossless body, $H$ is the mode sum~\eqref{eq:modesum}, which fixes the sign of $\im H$ at every complex frequency off the real axis:
\begin{equation}\label{eq:ImH}
\im H (\omega) = \im (\omega^2) \bigg[c_\infty+\sum_{n\ge1}\frac{c_n}{\left|\Omega_n^2-\omega^2\right|^2}\bigg]\;.
\end{equation}
The bracket is non-negative, so every lossless body satisfies
\begin{equation}
\label{eq:H_inequality}
\frac{\im H(\omega)}{\im \omega^2}\geq0 .
\end{equation}
A QNM can lie at $\omega$ only if $H_{\rm out}(\omega)$ satisfies Eq.~\eqref{eq:H_inequality}.
Figure~\ref{fig:qnm} (left) shows the frequencies this excludes for $\Rb=1.25\,\Rs$ (dark gray), together with the axial quadrupolar modes of a uniform-density star of that radius, all of which lie outside.\footnote{At this radius the central pressure of the star exceeds its energy density, so it serves as an exactly lossless example rather than a model of physical matter.}
The excluded region lies at low frequency and large damping and depends only on $\Rb/\Rs$.
It approaches the real axis only as $\omega\to0$.
On the real axis the outgoing solution carries energy outward, so $\im H_{\rm out}<0$ for $\re\omega>0$ at every radius, and weakly damped modes are always allowed.

\subsection{Love numbers improve the QNM bounds}
\label{sec:qnmlove}

The bounds in Sec.~\ref{sec:QNM_prop_a} apply to any lossless body, depending only on its mass and radius. 
The Love numbers are additional body-specific data from the low-frequency expansion of $H$.
Near zero frequency, the mode sum~\eqref{eq:modesum} reads
\begin{equation}\label{eq:Hexp}
    H=-c+H_2\omega^2+H_4\omega^4+\dots\;,
\end{equation}
where
\begin{equation}
\label{eq:Gdef}
    H_2\equiv\left.\frac{dH}{d\omega^2}\right|_{\omega=0}=c_\infty+\sum_n\frac{c_n}{\Omega_n^4}\ge0\;.
\end{equation}
Inverting Eq.~\eqref{eq:gmap} at zero frequency and at $\mathcal{O}(\omega^2)$ gives the first two coefficients:
\begin{equation}\label{eq:cG}
    c=\left.\frac{P^{(0)\prime}+\lambda_0\,Q^{(0)\prime}}{P^{(0)}+\lambda_0\,Q^{(0)}}\right|_{R_b}\,,\qquad \qquad H_2=\left.\frac{(2\ell+1)\,g_2}{\big(P^{(0)}+\lambda_0\,Q^{(0)}\big)^2}\right|_{R_b}\,,
\end{equation}
where $g_2=\lambda_2-\lambda_2^{\rm frozen}(\lambda_0)$ is defined in Eq.~\eqref{eq:g2inv} and does not depend on the scheme in which $\lambda_2$ is quoted.
The static Love number fixes $c$, and the two Love numbers together fix the slope $H_2$.
Each subsequent coefficient $H_{2n}$ in Eq.~\eqref{eq:Hexp} is fixed by $\lambda_0,\lambda_2,\dots,\lambda_{2n}$, narrowing the set of values that $H$ can take at every complex frequency and further restricting the allowed QNMs.

The Love numbers feed into the QNM bounds through a sequence of functions $f_1,\dots,f_N$ built from $H$. 
Each of these is a \emph{positive pole sum}, a function $f$ of the form
\begin{equation}\label{eq:possum}
f(\omega^2)=q_\infty+\sum_n q_n\,\frac{t_n}{t_n-\omega^2}\,,\qquad q_\infty,\,q_n\ge0\,,\quad t_n>0\,,
\end{equation}
for some poles $t_n$ in $\omega^2$ and weights $q_n$, $q_\infty$.
Following the logic of Sec.~\ref{sec:QNM_prop_a}, a positive pole sum obeys 
\begin{equation}\label{eq:pole_sum}
    \frac{\im f(\omega^2)}{\im\omega^2}\ge0.
\end{equation}

To generate positive pole sums we define the map 
\begin{equation}
    S[h]\equiv\frac{h-h(0)}{\omega^2}\,,
\end{equation}
which subtracts the static value of a function $h$ and divides by $\omega^2$.
Acting with the map $S$ on the impedance $H$ gives the first positive pole sum, 
\begin{equation}
    f_1\equiv S[H]=\frac{H+c}{\omega^2}=c_\infty+\sum_n\frac{c_n}{\Omega_n^4}\,\frac{\Omega_n^2}{\Omega_n^2-\omega^2}\;.
    \label{eq:improved_mode_sum}
\end{equation}
Its poles $t_n=\Omega_n^2$ are those of $H$ in Eq.~\eqref{eq:modesum}, and its weights are $q_n=c_n/\Omega_n^4$ and $q_\infty=c_\infty$. 
Evaluating $f_1$ at $H=H_{\rm out}$ requires only the constant $c=-H(0)$, which $\lambda_0$ fixes by Eq.~\eqref{eq:cG}.

This can be repeated with one further ingredient: the inversion $f\to-1/f$, which preserves the sign of $\im f/\im\omega^2$ since $\im(-1/f)=\im f/|f|^2$. 
If $f$ is a positive pole sum, then so is  
\begin{equation}\label{eq:step}
S\left[-\frac{1}{f}\right]=\frac{1}{\omega^2}\bigg(\frac{1}{f(0)}-\frac{1}{f}\bigg)\,,
\end{equation}
as follows from Stieltjes' continued-fraction expansion of $f$~\cite{Stieltjes:1894,Akhiezer:1965}.
Applying this step to $f_1$ gives a new positive pole sum with its own poles and weights:
\begin{equation}
    f_2=\frac{1}{\omega^2}\bigg(\frac{1}{H_2}-\frac{1}{f_1}\bigg)\,.
\end{equation}
This requires $f_1(0)=H_2$ and hence $\lambda_2$.
Applying the step again gives
\begin{equation}
    f_3=\frac{1}{\omega^2}\bigg(\frac{1}{f_2(0)}-\frac{1}{f_2}\bigg)\,.
\end{equation}
This requires $f_2(0)=H_4/H_2^2$ and hence $\lambda_4$.
Each further step requires one more Love number, which enters through the constant $f_k(0)$.
A lossless body with Love numbers $\lambda_0,\lambda_2,\dots,\lambda_{2N-2}$ therefore satisfies Eq.~\eqref{eq:H_inequality} and 
\begin{equation}\label{eq:qnmN}
\frac{\im f_k(\omega^2)}{\im\omega^2}\ \ge\ 0\,,\qquad k=1,2,\dots,N\,.
\end{equation}
For $k=1$ this is 
\begin{equation}
\frac{1}{\operatorname{Im}\omega^2}\,\operatorname{Im}\frac{H(\omega)+c}{\omega^2}\ \geq 0\;.
\label{eq:qnm1}
\end{equation}
For $k=2$, writing $\im f_2$ in terms of $f_1=(H+c)/\omega^2$ gives
\begin{equation}
    \left|\frac{H(\omega)+c}{\omega^2}\right|^2\ \le\ H_2\,\frac{\operatorname{Im}H(\omega)}{\operatorname{Im}\omega^2}\;.
    \label{eq:qnm2}
\end{equation}
A QNM can therefore lie at $\omega$ only if Eq.~\eqref{eq:qnmN} holds with $H$ replaced by $H_{\rm out}(\omega)$.

\begin{figure}[t]
\centering
\raisebox{-0.5\height}{\includegraphics[width=0.45\textwidth]{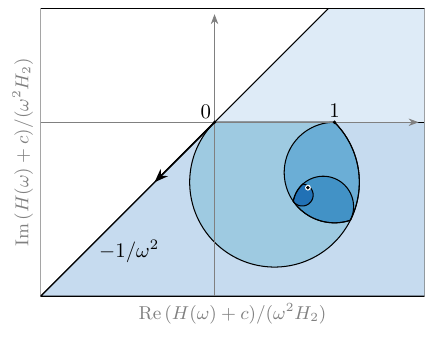}}%
\hfill
\raisebox{-0.5\height}{\includegraphics[width=0.55\textwidth]{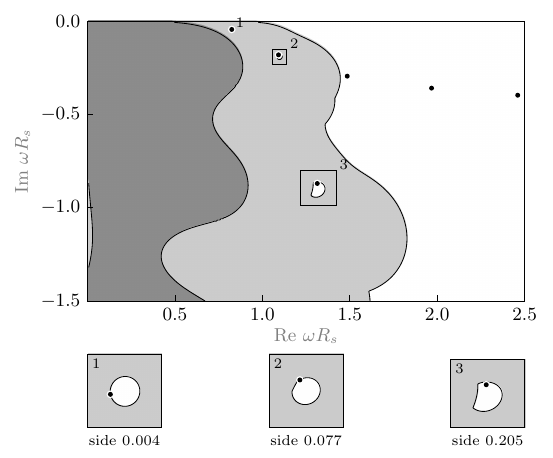}}
\caption{LEFT: Allowed values of $(H(\omega)+c)/\omega^2$, in units of $H_2$, at the fixed frequency $(\omega\Rs)^2=1-i$, for a lossless body with known first $N$ Love numbers $\lambda_0,\lambda_2,\dots,\lambda_{2N-2}$, for $N=0$ to $5$.
The shades add one Love number at a time, from light to dark.
With no Love numbers, the mode sum allows the half-plane bounded by the line through $0$ in the direction of $-1/\omega^2$ (arrow).
Adding $\lambda_0$ leaves the wedge between the positive real axis and the arrow.
Adding $\lambda_2$ leaves the part of the disk through $0$ and $1$ cut off by the chord joining them.
Adding $\lambda_4$, $\lambda_6$ and $\lambda_8$ leaves successively smaller disks, each from one further condition of Eq.~\eqref{eq:qnmN}.
A QNM at $\omega$ requires the exterior value $(H_{\rm out}(\omega)+c)/\omega^2$ to lie in the allowed set.
The regions for $N\geq3$ depend on the body; they are shown for the uniform-density star of Fig.~\ref{fig:qnm} (left), with $\Rb=1.25\,\Rs$, and the dot is the star's own value of $(H+c)/\omega^2$.
RIGHT: Frequencies excluded as QNMs of the same star, given its first seven Love numbers, $\lambda_0$ to $\lambda_{12}$.
Dark: excluded by the radius alone, the same region as the dark region of Fig.~\ref{fig:qnm} (left); light: additionally excluded by the seven Love numbers; white: allowed.
The dots are the star's QNMs.
Boxes 1 to 3, enlarged below, have widths $0.004$, $0.077$ and $0.205$ in units of $1/\Rs$, as labeled.}
\label{fig:qnmN}
\end{figure}

Figure~\ref{fig:qnmN} (left) shows the allowed set at one frequency.
Equation~\eqref{eq:H_inequality} gives the half-plane and Eq.~\eqref{eq:qnm1} the wedge.
Each further condition of Eq.~\eqref{eq:qnmN} for $k\geq2$ is a disk in the plane of $(H+c)/\omega^2$, with center and radius fixed by $\omega^2$ and the constants $f_1(0),\dots,f_{k-1}(0)$. 
The allowed set is the intersection of all the conditions, so it shrinks with each Love number.

Unlike the region of Sec.~\ref{sec:QNM_prop_a}, the regions excluded here depend on the body through its Love numbers.
To find where QNMs can lie we sweep over complex $\omega$ and apply at each frequency the test of Fig.~\ref{fig:qnmN} (left): if the exterior value $(H_{\rm out}(\omega)+c)/\omega^2$ falls outside the allowed set, the frequency is excluded (gray). 
Figure~\ref{fig:qnm} (left) shows the excluded frequencies for a uniform-density star of radius $\Rb=1.25\,\Rs$, whose first two Love numbers fix $c$ and $H_2$: the Love numbers enlarge the region excluded by the radius alone, and the least-damped mode of the star lies just outside the edge of the enlarged region. 
Figure~\ref{fig:qnmN} (right) shows the same map given the star's first seven Love numbers: the excluded region now surrounds the three lowest-frequency modes, leaving a small island around each, with the mode near its edge.

The damping rate $\gamma_f=-\im\omega_f$ of the $f$-mode is small compared with its frequency.
The QNM conditions derived here apply to it provided the star is lossless for polar tides, which we checked numerically (see Sec.~\ref{sec:bodies}). 
At the real part of the mode frequency, each condition gives an upper bound on the damping rate of the $f$-mode. 
For the 22 stars of Table~\ref{tab:stars} used in Fig.~\ref{fig:love}, the actual damping rate is $1$--$4\%$ of the bound from Eq.~\eqref{eq:H_inequality}, which uses the radius alone.
It is $34$--$81\%$ of the bound from Eq.~\eqref{eq:qnm1}, which adds $\lambda_0$, and $92.6$--$99.95\%$ of the bound from Eq.~\eqref{eq:qnm2}, which adds $\lambda_2$.

\subsection{Lower bounds on Love numbers from quasinormal modes}
\label{sec:modesLove}

The conditions of Sec.~\ref{sec:qnmlove} can be applied in the other direction.
At a QNM frequency $\omega$, the body's impedance equals the exterior's, $H(\omega)=H_{\rm out}(\omega)$, and the right-hand side is a known number once the complex frequency of the mode is known.
Each condition in Eq.~\eqref{eq:qnmN} then constrains the Love numbers rather than the frequency.
Condition $k$ contains one Love number that the lower conditions do not, and it bounds that Love number from below given the lower ones.
For $k=1$, Eq.~\eqref{eq:qnm1} gives $c\le|\omega|^4\,\im(H_{\rm out}/\omega^2)/\im\omega^2$, which through Eq.~\eqref{eq:lam0ofc} is a lower bound on $\lambda_0$ that needs only the mass, the radius and the mode.
For $k=2$, Eq.~\eqref{eq:qnm2} evaluated at the $f$-mode frequency $\omega_f$ gives a lower bound on the slope,
\begin{equation}\label{eq:floor}
    H_2\ \ge\ \left|\frac{H_{\rm out}(\omega_f)+c}{\omega_f^2}\right|^2\,\frac{\operatorname{Im}\omega_f^2}{\operatorname{Im}H_{\rm out}(\omega_f)}\;.
\end{equation}
Through Eq.~\eqref{eq:cG} this is a lower bound on $g_2$, and hence on $\lambda_2$ in any scheme, given the body's mass, radius, static Love number, and the complex frequency of its fundamental mode.

\begin{figure}[t]
\centering
\includegraphics[width=0.55\textwidth]{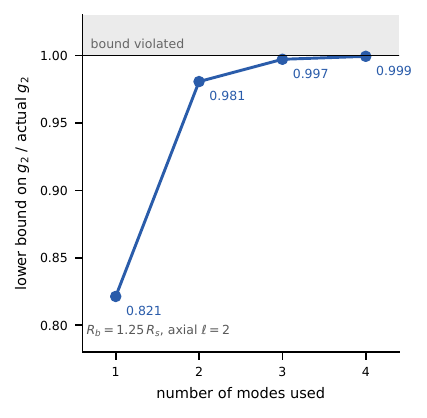}
\caption{The lower bound on $g_2$ given $\lambda_0$ and the least-damped axial QNMs of the uniform-density star of Fig.~\ref{fig:qnm} (left), divided by the actual value, versus the number of modes used.
One mode gives Eq.~\eqref{eq:floor}, and each further mode adds the condition $H=H_{\rm out}$ at its frequency.}
\label{fig:modesfloor}
\end{figure}

Figure~\ref{fig:qnm} (right) shows that this bound lies within 9\% of the actual $\lambda_2$ at $\mu=\re\omega_f$ for every star of Fig.~\ref{fig:love}. 
The bound is nearly saturated because the damping of the fundamental mode fixes the weight $\rho_1$ of the nearby pole in the mode sum~\eqref{eq:gsum} and almost nothing else.
The lower bound on $g_2$ that follows from Eq.~\eqref{eq:floor} is therefore approximately the share $\rho_1/\omega_1^4$ of $g_2$ carried by that pole, and for stars this pole carries most of $g_2$ (Sec.~\ref{sec:dynamical}).
In the Newtonian limit this share is proportional to the quadrupole damping rate of the mode~\cite{Thorne:1980ru}, and Eq.~\eqref{eq:floor} becomes $\lambda_2\ge90\,\gamma_f/(\re\omega_f)^8$.
Applied to relativistic stars, this Newtonian form fails for every star (Fig.~\ref{fig:qnm}, right), while Eq.~\eqref{eq:floor} holds.
Equation~\eqref{eq:floor} holds for every equation of state, whereas the known relations between the $f$-mode and the Love numbers are fits to stars computed for particular equations of state~\cite{Chan:2014kua,Saes:2025jvr}.

Figure~\ref{fig:modesfloor} shows that knowing further modes raises the floor.
Each known mode fixes the value of $(H+c)/\omega^2$ at its frequency.
To find the smallest $H_2$ compatible with these values, we write $(H+c)/\omega^2$ as the pole sum~\eqref{eq:improved_mode_sum} with poles on a fine grid of frequencies, and minimize the sum of the weights, which is $H_2$, over non-negative weights that reproduce these values (a linear program).

\section{Discussion}
\label{sec:discussion}
The surface impedance is fixed by the interior alone, so bounds on it constrain the body without the long-range effects of the exterior. Moreover, it fixes the response to arbitrarily fast tides, which is negative in general relativity.
Bounds on the body's Love numbers and QNMs follow through the impedance from either of two general properties: that the body is lossless, or that it is passive. 
Because the bounds are imposed at the surface, before the dressing by the exterior field that produces running, the bounds on Love numbers hold in every scheme and at every scale.
Furthermore, knowledge of the first few Love numbers sharpens the bounds on the QNMs, and vice versa, because both are fixed by the same surface impedance: the Love numbers are the low-frequency coefficients of the response it determines, and the QNMs are the frequencies at which it matches the outgoing exterior. 
In practice, measured values of the mass, radius, Love numbers and fundamental mode that violate Eq.~\eqref{eq:floor} would rule out a lossless body with a Schwarzschild exterior.

The most direct extension of our bounds is to rotating bodies, which are phenomenologically relevant for LISA observations. 
The lossless response of a rotating body contains terms odd in $\omega$~\cite{Steinhoff:2021dsn}. 
The impedance is then no longer a function of $\omega^2$ alone, so the mode sum~\eqref{eq:modesum} must be restated. 
Additionally, a rotating body can amplify waves through a phenomenon known as \emph{superradiance}~\cite{Zeldovich:1971ffh,Starobinsky:1974nkd,Brito:2015oca}. 
The upper bound on the static Love numbers of rotating bodies therefore requires a restatement of passivity. 
Another direction that would connect these results to phenomenology is to bound the tidal response at the frequencies of the inspiral, which waveform models need. 
Our bounds constrain the low-frequency coefficients of the response and the positions of the modes, whereas the inspiral drives the body at tidal frequencies that reach a sizeable fraction of the fundamental mode before merger.

Our two-sided dynamical bounds require a lossless body.
Including dissipation changes the response in two ways: odd powers of $\omega$ appear in its low-frequency expansion with the absorptive Love numbers as coefficients, and the discrete mode sum becomes a measure with a continuous part that can extend to zero frequency.
The Love numbers are then no longer the moments of a positive measure with a gap below the lowest mode, which is what the lossless bounds use. 
This is the same obstruction to bounds that Ref.~\cite{Apostolidis:2026mvb} finds.
Bounding the dynamical Love numbers of a weakly dissipative body would therefore require new assumptions.

It would also be interesting to derive our bounds from scattering amplitudes, connecting more closely with the positivity-bound literature.
The partial-wave amplitude is a M\"obius image of $H$ with coefficients fixed by the exterior, Eq.~\eqref{eq:Sell}, so the bounds are positivity conditions on the amplitude once the known exterior is removed.
This suggests a route through the worldline EFT that avoids the forward-limit singularity of the full amplitude.

\begin{acknowledgments}
We thank Paolo Pani for suggesting the gravastar tests and Anna Wolz for feedback on the draft. D.~M. and M.~S. are supported by the US Department of Energy under award number DE-SC0024224, the Sloan Foundation, and the Mani L. Bhaumik Institute for Theoretical Physics.
This project used Claude, Mathematica, and computational and storage services associated with the Hoffman2 Cluster, which is operated by the UCLA Office of Advanced Research Computing's Research Technology Group.
\end{acknowledgments}

\appendix
\section{Stability}
\label{app:stability}
A body is stable if no perturbation of it grows exponentially in time.
We use two consequences of stability for a lossless body: the static Love number obeys the lower bound of the window~\eqref{eq:window}, and the poles of Eq.~\eqref{eq:modesum} lie at $\Omega_n^2>0$.
Here we derive both, and show that a passive body is stable without further assumption.

We first show that a lossless body with $\lambda_0<\lamStiff$ has a mode that grows exponentially in time.
We look for it on the positive imaginary axis, where it satisfies the QNM condition~\eqref{eq:qnmcond}, grows as $e^{|\omega|t}$, and the outgoing solution decays at infinity.
On this axis both $H$ and $H_{\rm out}$ are real.
Along it, the difference $H-H_{\rm out}$ is positive at $\omega=0$ and negative at large $|\omega|$, as we check below.
If $H$ has no pole on the axis, the difference therefore vanishes somewhere in between, and that zero is a growing mode.

For a lossless body, a pole of $H$ on this axis also gives a growing mode.
On this axis $\omega^2=-|\omega|^2$, so $H$, which increases with $\omega^2$, decreases as $|\omega|$ grows.
Just past the pole $H$ is therefore large and positive, and so is $H-H_{\rm out}$.
Farther out, at large $|\omega|$, the difference is negative, so it vanishes in between.
This is why the poles in Eq.~\eqref{eq:modesum} lie at $\Omega_n^2>0$.
It also means that a lossless body with $\lambda_0<\lamStiff$ has a growing mode whether or not $H$ has a pole on the axis.

It remains to verify the two signs used above: $H-H_{\rm out}$ is positive at $\omega=0$ when $\lambda_0<\lamStiff$, and negative at large $|\omega|$.
At $\omega=0$, $H=-c$, while $H_{\rm out}=-Q^{(0)\prime}/Q^{(0)}|_{\Rb}$ because the outgoing solution becomes the static solution $Q^{(0)}$, which decays at infinity.
By Eq.~\eqref{eq:lam0ofc}, $H$ is the larger exactly when $\lambda_0<\lamStiff$.
At large $|\omega|$, $H_{\rm out}\simeq-i\omega=|\omega|$ grows without bound.
In contrast, $H\simeq i\omega=-|\omega|$ for finite signal speed, and Eq.~\eqref{eq:modesum} bounds $H$ from above for a lossless body, so $H$ is the smaller.
This completes the proof that a stable lossless body has $\lambda_0\ge\lamStiff$.

A passive body has no growing mode, because $H=H_{\rm out}$ has no solution with $\im\omega>0$.
There Eq.~\eqref{eq:PR} gives $\im(H/\omega)\ge0$.
In contrast, multiplying Eq.~\eqref{eq:master} by $\bar u_{\rm out}$ and integrating from $\Rb$ to infinity gives $H_{\rm out}=A-\omega^2B$.
Here $A$ and $B$ are the integrals over $r_*$ of $|u_{\rm out}'|^2+V|u_{\rm out}|^2$ and of $|u_{\rm out}|^2$, each divided by $|u_{\rm out}(\Rb)|^2$, so both are positive.
Hence $\im(H_{\rm out}/\omega)=-(\im\omega/|\omega|^2)(A+|\omega|^2B)<0$.
Nor does it need the argument above for the lower bound: it has $c\ge0$, and Eq.~\eqref{eq:lam0ofc} maps every $c\ge0$ into the window (Sec.~\ref{sec:window}).

\section{The energy at the surface}
\label{app:energy}
The energy near the body is not unique.
Adding a surface term $\frac12\beta\,u^2|_{\Rb}$ to Eq.~\eqref{eq:energy}, with $\beta$ a constant, changes the power into the body to $p=\dot u\,(u'-\beta u)$, which replaces $H$ by $H+\beta$.
For a black hole the choice is fixed.
At the horizon the field is purely ingoing, so $u'=\dot u$ and $p=\dot u\,(\dot u-\beta u)$, and since the horizon absorbs at every instant~\cite{Hawking:1971tu,Hawking:1972hy}, $p\ge0$ for every field only if $\beta=0$.
For a surface above the horizon we assume $\beta=0$ throughout.

A different $\beta$ does not change which bodies are lossless, or any bound for lossless bodies, since a constant added to $H$ keeps the form \eqref{eq:modesum}.
It also leaves the average power at real frequencies unchanged, since $\langle u\dot u\rangle=0$ there.
It changes the bounds for passive bodies, because the condition \eqref{eq:staticslope} becomes $c\ge\beta$.
For $\beta>Q^{(0)\prime}/Q^{(0)}|_{\Rb}$ the upper bound of the window \eqref{eq:window} becomes the value of Eq.~\eqref{eq:lam0ofc} at $c=\beta$, and for smaller $\beta$ there is none; Eq.~\eqref{eq:combined} changes accordingly.
Near the horizon this shifts $k_2^{E,\rm free}$ in Eq.~\eqref{eq:windowhorizon} by $-\tfrac{8}{15}\beta\Rs$, which exceeds $\tfrac{72}{35}\delta$ unless $\beta$ vanishes at least as fast as $\delta$.
If the surface term has a fixed size in the frame of the surface, the redshift gives $\beta\propto\sqrt\delta$, and the upper bound then closes as $\sqrt\delta$ rather than linearly.

\section{Data for the stars}
\label{app:stars}

Table~\ref{tab:stars} lists the data for the 22 stars of Figs.~\ref{fig:love} and~\ref{fig:qnm} (right) and of Sec.~\ref{sec:qnmlove}.
The table omits the response to arbitrarily fast tides, since it depends only on the compactness: $\ginf=\lamStiff(\Rb)$, Eq.~\eqref{eq:ginf}.
Our numerics reproduce the black-hole values of $\lambda_2$~\cite{Combaluzier--Szteinsznaider:2025eoc} to $3\times10^{-8}$, the static Love numbers that Ref.~\cite{Binnington:2009bb} tabulates for polytropes defined with the energy density instead of the rest-mass density, and the least-damped axial mode of the uniform-density star in Table~\ref{tab:uniform}~\cite{Andrade:1999mj}.
For polar tides our $\lambda_2$ agrees exactly with the expression of Ref.~\cite{Apostolidis:2026qsg} for a general body.

Table~\ref{tab:uniform} lists the data for the uniform-density star of Figs.~\ref{fig:qnm} (left), \ref{fig:qnmN} and~\ref{fig:modesfloor}, for axial tides with $\ell=2$.
The dot in Fig.~\ref{fig:qnmN} (left) is the star's own value of $(H+c)/\omega^2$ at $(\omega\Rs)^2=1-i$, which is $0.778721-0.545068\,i$ in units of $H_2$.

\begin{table}[t]
\centering
\small
\begin{tabular}{lcccccccc}
\toprule
EOS & $C$ & $M/M_\odot$ & $k_2$ & $g_2/\Rb^7$ & $\lambda_2/\Rb^7$ & $\omega_1\Rb$ & $\re\omega_f\Rb$ & $10^5\,\gamma_f\Rb$ \\
\midrule
$n=0.5$ & 0.05 & -- & 0.323807 & 70.4679 & 70.5514 & 0.235982 & 0.233992 & 0.758710 \\
$n=0.5$ & 0.10 & -- & 0.224908 & 24.6037 & 24.3868 & 0.335380 & 0.330561 & 4.77948 \\
$n=0.5$ & 0.15 & -- & 0.148811 & 11.1717 & 10.8960 & 0.409630 & 0.402253 & 12.3347 \\
$n=0.5$ & 0.20 & -- & 0.0920820 & 5.51651 & 5.29419 & 0.467213 & 0.458114 & 21.1272 \\
$n=0.5$ & 0.25 & -- & 0.0514871 & 2.76730 & 2.62350 & 0.509320 & 0.499769 & 26.9514 \\
$n=0.5$ & 0.30 & -- & 0.0239166 & 1.33050 & 1.23509 & 0.533546 & 0.525252 & 24.8979 \\
\midrule
$n=1$ & 0.05 & -- & 0.179190 & 28.3972 & 28.1375 & 0.277238 & 0.275424 & 1.12983 \\
$n=1$ & 0.10 & -- & 0.117101 & 9.44134 & 9.13165 & 0.393540 & 0.389343 & 6.89141 \\
$n=1$ & 0.15 & -- & 0.0707513 & 3.99596 & 3.73211 & 0.480138 & 0.474140 & 16.9596 \\
$n=1$ & 0.20 & -- & 0.0372906 & 1.75005 & 1.56021 & 0.547678 & 0.541126 & 26.6487 \\
$n=1$ & 0.21 & -- & 0.0318626 & 1.46892 & 1.29310 & 0.559335 & 0.552886 & 27.8380 \\
\midrule
$n=1.5$ & 0.05 & -- & 0.0917059 & 10.3522 & 10.1106 & 0.329596 & 0.328007 & 1.66924 \\
$n=1.5$ & 0.075 & -- & 0.0709743 & 5.39568 & 5.15981 & 0.405361 & 0.402799 & 4.81838 \\
$n=1.5$ & 0.10 & -- & 0.0531912 & 3.09285 & 2.87677 & 0.469473 & 0.466087 & 9.62389 \\
$n=1.5$ & 0.13 & -- & 0.0352425 & 1.64245 & 1.46037 & 0.536726 & 0.532738 & 16.5994 \\
\midrule
SLy & 0.05 & 0.4224 & 0.0748957 & 7.44285 & 7.22408 & 0.352869 & 0.351252 & 2.07995 \\
SLy & 0.10 & 0.8078 & 0.104687 & 7.97526 & 7.67474 & 0.405943 & 0.401787 & 7.51186 \\
SLy & 0.15 & 1.1991 & 0.0898142 & 5.51705 & 5.23074 & 0.458025 & 0.451533 & 15.7061 \\
SLy & 0.1766 & 1.4000 & 0.0761666 & 4.30595 & 4.04958 & 0.482206 & 0.474753 & 20.3339 \\
SLy & 0.20 & 1.5681 & 0.0635330 & 3.40248 & 3.17924 & 0.501005 & 0.492936 & 23.9794 \\
SLy & 0.25 & 1.8781 & 0.0381787 & 1.94720 & 1.79595 & 0.532483 & 0.524034 & 28.2872 \\
SLy & 0.30 & 2.0473 & 0.0178675 & 0.997925 & 0.889394 & 0.549093 & 0.541981 & 24.1680 \\
\bottomrule
\end{tabular}
\caption{The stars used in this paper, for polar tides with $\ell=2$: compactness $C=M/\Rb$, mass (for SLy), static Love number $k_2=\lambda_0/(12\Rb^5)$, the body's part $g_2$ of the leading dynamical Love number, $\lambda_2$ in the asymptotic scheme at $\mu=\re\omega_f$, the lowest pole $\omega_1$ of $g$, Eq.~\eqref{eq:gsum}, and the frequency and damping rate $\gamma_f=-\im\omega_f$ of the fundamental mode.}
\label{tab:stars}
\end{table}

\begin{table}[t]
\centering
\small
\begin{tabular}{ccccccc}
\toprule
$c$ & $H_2$ & $H_4$ & $H_6$ & $H_8$ & $H_{10}$ & $H_{12}$ \\
\midrule
0.747180 & 0.669191 & 0.280491 & 0.213409 & 0.187120 & 0.172456 & 0.162153 \\
\bottomrule
\end{tabular}

\vspace{1em}
\begin{tabular}{cccc}
\toprule
\multicolumn{4}{c}{QNMs $\omega\Rs$} \\
\midrule
$0.824693-0.0439846\,i$ & $1.09214-0.180232\,i$ & $1.31397-0.870147\,i$ & $1.37412-1.85423\,i$ \\
$1.48538-0.294539\,i$ & $1.96783-0.358855\,i$ & $2.46064-0.396784\,i$ & $2.95449-0.426262\,i$ \\
\bottomrule
\end{tabular}
\caption{The uniform-density star with $\Rb=1.25\,\Rs$, for axial tides with $\ell=2$, in units $\Rs=1$. Top: the stiffness $c=-H(0)$ and the coefficients $H_{2n}$ of the low-frequency expansion~\eqref{eq:Hexp}, which are fixed by the Love numbers $\lambda_0,\dots,\lambda_{2n}$. Bottom: the QNMs shown in Figs.~\ref{fig:qnm} (left) and~\ref{fig:qnmN} (right), ordered by frequency.}
\label{tab:uniform}
\end{table}

\clearpage
\bibliography{refs}
\end{document}